\documentclass[11pt]{article}
\usepackage{geometry} 
\usepackage{amsmath}
\usepackage{amssymb}
\usepackage{amsbsy}
\usepackage{ascii}
\usepackage[margin=30pt]{caption}
\usepackage{bm}
\usepackage{bbold}
\usepackage{mathrsfs}
\usepackage{graphicx,subfigure} 
\usepackage{wrapfig}
\usepackage{ifthen}
\usepackage{float}
\usepackage{epstopdf}
\usepackage[pdftex]{color}
\usepackage[all]{xy}

\title{ Quantum Operations: technical or fundamental challenge?}
\author{Bogdan Mielnik}
\date{}
\begin{document}
\maketitle

{\small
\noindent \hspace{1cm} Depto. de F\'isica, Centro de Investigaci\'on y de Estudios Avanzados del IPN,

\hspace{.4cm} A.P. 14-740, M\'exico, DF 07000, Mexico. e-mail: bogdan@fis.cinvestav.mx \\
}

{\small
\hspace{.4cm} Keywords: quantum control, soft operations.
}
\begin{abstract}
A class of unitary operations generated by  idealized, semiclassical fields is studied. The  
operations implemented by sharp potential kicks are  revisited and the 
possibility of performing them by softly varying external fields is examined. 
 The possibility of using the ion  traps as 'operation factories' 
transforming  quantum states is discussed. The non-perturbative algorithms indicate  
 that the results of abstract $\delta$-pulses of oscillator potentials  can become real. 
Some of them, if empirically achieved,  could be essential to examine certain 
atypical quantum ideas.  In particular, simple    
dynamical manipulations  might contribute to the Aharonov-Bohm criticism  of the   
time-energy uncertainty principle, and some others,  to verify the existence of fundamental precision 
limits of the position measurements or the reality of `non-commutative geometries'.
\end{abstract}

{\small \noindent \hspace{.8cm} PACS numbers: 03.65.Fd, 03.67.Ta}

\section{Introduction}

One of  limitations of the present day quantum theories is the   
``passive'' evolution picture in which the physical systems (i.e., some parts of the 
 universe) evolve under the influence of "the rest", typically represented by  some given (if not stationary) 
external conditions, and the role of active state manipulations is  reduced to the choice of   initial or boundary conditions. However, if the physical theories were at all created it is only because the physical systems can be actively manipulated, by changing the  
external conditions  and performing experiments.

In spite of the limited use of dynamical manipulations, the present day quantum theories  
show some spectacular achievements. However, 
the concepts applied and  questions asked are somewhat repetitive. What dominates are some  
idealized scenarios with 
pure states described always by vectors in the linear (Hilbert) spaces and the
time evolution (in absence of dissipation) obeying always the linear, unitary operations.    
The picture persists in the description  of composite systems and quantum field theories (QFT)
where the states are always represented by tensor product spaces - in a    
desire to conserve the linearity of basic laws at the cost of 
multiplying the number of variables.  

Some doubts, though, persist. Can indeed all 'gedanken states', represented by vectors in Hilbert spaces be physically created \cite{Primas}? Moreover, can all unitary operations be achieved (or at least approximated) by physical evolution? In the recent research a lot of attention is dedicated to the finite-dimensional spin states (qubits) with hopes to develop quantum computers. However, can  the problem be indeed  
exhausted by tensor products with complex coefficients?

One of main troubles in checking  the 'obligatory beliefs' of quantum  theories,  are the perturbative complications as well as the difficulties of extrapolating toward 'the great' and toward 'the small' \cite{Penrose}.  What could help are the exact solutions, though they seldom exist. Yet, in certain areas, there appear some windows in perturbative clouds. This happened  in the bound state manipulation \cite{Haroche, PhToday} (Nobel Prize 2012 for S. Haroche and D. Wineland),  in macroscopic superpositions of Leggett \cite{Leggett}, then  in works on "quantum tomography" \cite{Mancini, MaMaTo,TomHilbert, Marmo};  last not least, in the duality links  between the quantum quark and classical string dynamics \cite{Malda, Holo2}. The purpose of this study is to show that even the well known and  modest case of classical-quantum duality, for particles in time dependent quadratic potentials, 
has still some unexplored consequences.

The article is organized as follows. In  Sec.2. the particle behavior  in 1-dimensional quadratic, time dependent potentials $V(q,t)=\beta(t)\frac{q^2}{2}$ is classified in most  elementary terms.  
Sec.3 reports briefly this classification analogues in the macroscopic world. In Sec.4 a class of idealized  evolution effects produced  by $\delta$-kicks of the quadratic potentials is presented, and Sec.5 outlines their  optical equivalents. Their links with the arguments of Aharonov, Bohm et al. against the time-energy uncertainty principle. are reported in Sec.6. The next Secs.7-8 show how to design  the soft equivalents of the singular pulse operations. The  last Sec.9 discusses their possible fundamental implications.

\section{Quadratic Hamiltonians: the classical-quantum structures}

The quantum theories satisfy the {\em correspondence principle},  becoming classical in the 
macroscopic limit, as $\hbar \rightarrow 0$. However,  some mathematical 
aspects are shared by classical and quantum theories without the need of any limiting transition. 
The simplest cases occur for non-relativistic time dependent, quadratic Hamiltonians in 1D:

\begin{equation}
\label{n1}
H(t) = \frac{p^2}{2} + \beta(t)\frac{q^2}{2}
\end{equation}
where $q$ and $p$ are the canonical position and momentum variables (we put for simplicity the mass $m=1$). Below, we shall not consider any  higher space dimensions, nor the deeper Hilbert space problems 
(carefully reviewed by  Barry Simon \cite{Simon}). The only challenge 
we attend is  strictly combinatorial: how should one program  the oscillations of the $c$-number amplitude $\beta (t)$, to generate some useful quantum control operations? On purely mathematical level, the problem is elementary. In classical theory and for nonsingular $\beta (t)$ the canonical equations lead to the linear evolution 
transformations of the canonical variables, represented by the  $2\times2$ symplectic {\em evolution matrices} $u(t,t_0)$: 

\begin{equation}
\label{n2}
\begin{pmatrix} q(t) \\ p(t)\end{pmatrix}  = u(t,t_0) \begin{pmatrix} q(t_0) \\ p(t_0)\end{pmatrix}
\end{equation}
given by the matrix equations: 
\begin{equation}
\label{n3}
\frac{d}{dt}u(t,t_0) = \Lambda(t) u(t,t_0) \qquad \frac{d}{dt_0}u(t,t_0) = -u(t,t_0) \Lambda(t_0) 
\end{equation}
with:
\begin{equation}
\label{n4}
\Lambda(t)  = \begin{pmatrix} 0 & 1 \\ -\beta(t) & 0 \end{pmatrix}
\end{equation}
and: 

\begin{equation}
\label{n5}
u(t,\theta)u(\theta,t_0) = u(t,t_0) \qquad u(t_0,t_0) = \mathbb{1}
\end{equation}

In quantum theory, the corresponding evolution operators $U(t,t_0)$, are  given by:  
  
\begin{equation}
\label{n6}
i\frac{d}{dt}U(t,t_0) = H(t)U(t,t_0), \qquad i\frac{d}{dt_0}U(t,t_0) = -U(t,t_0)H(t_0)
\end{equation}
with: 

\begin{equation}
\label{n7}
U(t,\theta)U(\theta,t_0) = U(t,t_0) \qquad U(t_0,t_0) = 1
\end{equation}
defining the evolution of the observables $q$ and $p$  in Heisenberg's picture  
(the {\em "Heisenberg's trajectory"}),  given by the same symplectic matrices (3-5):

\begin{equation}
\label{n8}
U(t,t_0)^\dagger\begin{pmatrix} q \\ p\end{pmatrix} U(t,t_0) = u(t,t_0) \begin{pmatrix} q \\ p\end{pmatrix}
\end{equation}
though now $q, p$ are operators with $[q,p]=i$ (we put for simplicity $\hbar=1$). The  exact formal correspondence of the classical and quantum cases was explored by multiple research groups, quite frequently without knowing of each other. In all these approaches, a useful observation  is 
\bigskip
 
{\bf Proposition 1.} In absence of spin or any additional degrees of freedom, each unitary evolution operator $U(t,t_0)$ in $L^{2}(\mathbb{R})$ generated by the quadratic, time dependent Hamiltonians \eqref{n1} is   determined  up to a phase factor by the classical motion trajectories. 

{\bf Proof.} Instead of sophisticated arguments,  notice simply that if two unitary operators $U_1$ and $U_2$ produce the same transformation of the canonical variables i.e., $U^{\dagger}_1qU_1 =U^{\dagger}_2qU_2 $ and  $U^{\dagger}_1pU_1 =U^{\dagger}_2pU_2 $, then $U_1U^{\dagger}_2$ commutes with both 
$q$ and $p$. Hence, it commutes also with any  function  of $q$ and $p$, including their spectral projectors. Since in $L^2 (\mathbb{R})$ the 
functions of $q$ and $p$ generate an irreducible algebra, then $U_1U^{\dagger}_2$ must be a c- number and since it is unitary, it can be only a phase factor,  $U_1U^{\dagger}_2= \text{e}^{i\varphi} \Rightarrow U_1 =\text{e}^{i\varphi} U_2$ where $\varphi \in \mathbb{R}$ \cite{RS, Simon, BM77}. $\square$
\bigskip

Any two unitary operators which differ only by a c-number phase factor generate the same transformation of quantum states, so  we shall call them {\em equivalent} 
 and  write $U_1 \equiv U_2$ (the fact that they might wear different phase factors can be of interest for the 
linear representation theory \cite{MSWolf,Historical,DavInt} but does not affect the operations performed on physical states, the principal subject of our interest).  

The possibility of deducing the quantum state evolution  from the transformations of the canonical variables (\ref{n8})  permits one to program ample families of the exact classical/quantum control operations. For their classification, the algebraic types of  matrices (\ref{n2}-\ref{n8}) are quite essential. Since every  evolution matrix $u= u(t,t_0)$ is symplectic  (Det$u =1$)  its algebraic type is defined just by one invariant Tr$u$. The characteristic equation: 

\begin{equation}
\label{characteq}
\operatorname{D}(\lambda) = \operatorname{Det}(\lambda - u)= \lambda^2 - \operatorname{Tr}(u)  \lambda + 1 = 0
\end{equation}
has two roots $\lambda_\pm = \frac{1}{2}\operatorname{Tr}u \pm i \sqrt{\Delta}$, where $\Delta= 1-\frac{1}{4}(\operatorname{Tr}u)^2$, permitting to distinguish three types of  evolution matrices: 
\bigskip

({\bf I}) If $ |\operatorname{Tr}(u)|<2$, then $u$ has two complex  eigenvalues  $\lambda_\pm =\text{e}^{\pm i \sigma}$ where $0\neq \sigma \in\mathbb{R} $ 
\bigskip

({\bf II}) If  $ |\operatorname{Tr}(u)|=2$ then $u$ is in the threshold: $\lambda_+ =\lambda_- =\pm1$ 
\bigskip

({\bf III}) If $ |\operatorname{Tr}(u)|>2$,  then $u$ has two real  eigenvalues  $\lambda_\pm =\text{e}^{\pm  \sigma}$ where $0\neq \sigma \in\mathbb{R} $ 
\bigskip

 The classification turns specially relevant if the function $\beta (t)$ in (\ref{n1}) is periodic, $\beta (t+T)=\beta (t)$, defining a {\em Floquet process}.
The (crucial) {\em Floquet matrices}  $u(t_0+T, t_0)$  define then the repeated evolution incidents. One easily shows that their types do not depend on $t_0$. Choosing $t_0=0$ and denoting for simplicity $u(t)=u(t,0)$ one sees that the results of the evolution in the sequence of expanding intervals $[0,nT]$ are given by repetitions of $u(T)$, i.e., $u(nT) = u(T)^n$. Now, if $u(T)$ is in the class ({\bf I}) the evolution is oscillatory. The eigenvectors of $u(T)$ define a pair of  variables $A_\pm$ (generalizing the {\em creation} and {\em annihilation} operators) which for $t=nT$ perform just the phase rotations $U(t)^{\dagger}A_{\pm}U(t) = \text{e}^{\pm i\sigma t}A_\pm$. However, if $u(T)$ is in  the class ({\bf III}) then the equilibrium is lost: the eigenvectors of $u(T)$ define two real canonical variables $A_\pm$ which are multiplied by $\text{e}^{\pm\sigma t}$, ($0\neq\sigma\in\mathbb{R}$) i.e., endlessly squeezed or endlessly  amplified as $t=nT\rightarrow \infty$. In turn, the threshold cases ({\bf II}) offer some exceptional  manipulation techniques; their mechanism is  simple, but  consequences are not.

\section{The Microscopic models of  Macroscopic Universe} 

The properties  of (\ref{n1}) intrude also into the spectral problems of the time independent energy operators $H=-\frac{1}{2}\frac{d^2}{dx^2} + V(x)$ with arbitrary real potentials $V(x)$. Indeed, by looking for the real stationary solutions $\psi(x)$ of the Shrodinger equation:

\begin{equation}
\label{e9}
-\frac{1}{2}\frac{d^2\psi}{dx^2} + [ V(x) -E]\psi(x) =0 
\end{equation} 
for any real $E$, belonging or not to the spectrum of $H$, one sees that the pair of variables $\psi$, $\psi'=\frac{d\psi}{dx}$ is displaced along the x-axis according to the 1-st order matrix equation: 

\begin{equation}
\label{e10}
\frac{d}{dx}\begin{pmatrix} \psi \\ \psi '\end{pmatrix}  =  \begin{pmatrix} 0 & 1 \\ -\beta_E(x) & 0 \end{pmatrix} \begin{pmatrix} \psi \\ \psi '\end{pmatrix}
\end{equation}
where $\beta_E(x) =2 [E- V(x)]$. So, by reinterpreting the variable $x$ as $t$, $\beta_E(x)$ as $\beta_E(t)$, then  $\psi(x)$ and $\psi'(x)$ as $q(t)$ and $p=\frac{dq}{dt}$ respectively, one sees that the spectral problem (\ref{e9}) has a classical equivalent in form of the oscillator (\ref{n1})  with a time dependent elasticity coefficient $\beta_E(t)$ \cite{RS, Magnus}. The spectrum and the spectral gaps of quantum system (\ref{e9}) correspond exactly to the stability and instability (parametric resonance) zones  of the classical oscillator (\ref{n1}) with some consequences for macroscopic phenomena. Curiously, certain  aspects of both pictures are opposite. What  in quantum case  was the synonym of stability (an energy eigenstate) corresponds to an exceptional, highly unstable orbit in the classical picture. This equivalence inspired Avron and Simon to describe the structure of Saturn rings by spectral bands and gaps of the Schrodinger's energy operator \cite{Avron} (an analogy worth attention, even if astronomers opted for different models). In a different scale, the abrupt changes of the classical trajectories (\ref{e10}) as $E$ evolves,  reflecting  the behavior of the Schrodinger's wave function (\ref{e9}), can be used in  an efficient  computer method to solve the spectral problems of the potential wells by observing the instabilities of classical trajectories  \cite{Classical}.

Some intriguing cases of quantum-classical  duality  appear in the relativistic cosmology, where certain variants of the Schroedinger equation are used to extrapolate the past  cosmos evolution near the Big-Bang \cite{Lidsey,Leach,Weinberg}.  If such cosmological reinterpretation were correct, then due to the exceptional character of the Schroedinger energy eigenstates,  it could also reveal some instability mechanisms of Avron-Simon type  in our reconstruction of the early universe (or its distant future), usually neglected in the cosmological 'slow roll ' models \cite{Weinberg} (see also discussions in \cite{Marco, Schulze}). While the cosmic parameters can be studied  (but not manipulated), our purpose below will be to discuss the parallel  chapters of quantum control, which can suggest the laboratory operations.

\section{The evolution  controlled by sharp pulses} 

 The exact solutions of  (\ref{n1}) were widely explored by  Lewis and Riesenfeld \cite{LeRie1, LeRie2, Leach}, then by Malkin, Man´ko, Trifonov \cite{MM1, MM2} in terms of {\em adiabatic invariants}, interrelated also with  an important techniques of {\em quantum tomography} \cite{Mancini, MaMaTo,Marmo}.  The mathematical algorithms, though elementary, are not immediate to apply (they require anyhow the solution of the 2-nd order differential eq.\eqref{e9}, complicated frequently by pertubative difficulties). 

 As independently observed, one of the  simplest ways to control the dynamical evolution consists 
in generating a closed evolution pattern ({\em evolution loop}) and 
then considering their perturbed or deformed versions. If the unperturbed evolution is represented by a certain family of unitary operators $U_0(t)$, then the perturbed evolution  operators split into $U(t)= U_0(t) W(t)$, where $U_0(t)$ represent the basic dynamical 
process and $W(t)$ is the correction (the evolution in the  {\em  interaction frame}). If now at some moment $t=T$ the basic evolution closes to a loop $U_0(T)\equiv 1$, then the full evolution reduces just to the pure deformation, $U(T)\equiv W(T)$ represented on Fig.1, in general, much easier to manipulate by the external fields. The most elementary evolution loops occur in the time independent oscillator potentials, leading e.g., to the {\em non-demolishing quantum  measurements} \cite{Thorne}, but do not exhaust the manipulation techniques.

\begin{figure}[htpb]
\centering
\includegraphics[width=5.5cm]{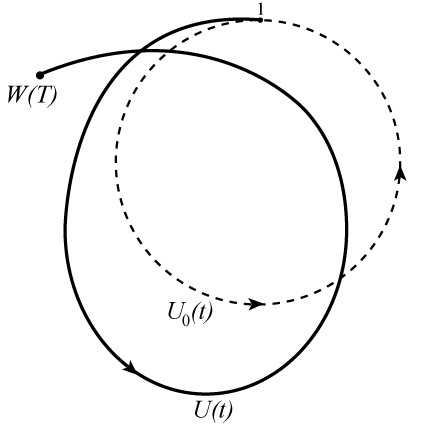}
\caption{The applications of an evolution loop. The basic and perturbed evolution processes are represented by $U_0(t)$ and $U(t)=U_0(t)W(t)$, where the 
$W(t)$ is the {\em evolution in the interaction frame}. If for some $T$, the $U_0(T)$ closes to the evolution loop, $U_0(T)\equiv  1$, then the whole process reduces to $W(T)$ alone, the {\em precession operator}, sensitive to  manipulation programs.} \label{fig1}
\end{figure}

The existence of  {\em non-adiabatic} loops generated by time dependent oscillator forces was 
noticed in 1970 by Malkin and Man'ko \cite{MMloop}, though without elaborating the operational 
consequences. The possibility of driving the quantum states by $\delta(t)$-pulses of the external 
fields was considered by Lamb Jr.\cite{Lamb}.  An extremely  simple class of exact though 
formal solutions of \eqref{n1} in  $L^2(\mathbb{R})$  was obtained  in \cite{BM77, Loops,Control,Dav11,
Dav12} by superposing two types of     
elementary operations:  the incidents of free evolution and the effects of the sharp pulses of  
oscillator potentials. Each  free evolution incident  in any interval $[t_0,t_1]$ produces the 
unitary evolution operator $e ^{-i\tau \frac{p^2}{2}}$,  
where $\tau= t_1-t_0$. In turn, the result of each sudden $\delta$-kick of the quadratic potential $V(q,t)=a\delta (t-t_0) \frac {q^2}{2}$  (where $a$ is the pulse amplitude), 
is most easily described by adopting the 
rectangular $\delta$-model defined by $\delta_\epsilon (t)= \frac{1}{\epsilon}$ in a narrow interval $[t_0, t_0 + \epsilon]$ and vanishing outside. The 
evolution in  $[t_0, t_0 + \epsilon]$ is then induced by the constant Hamiltonian $H_\epsilon = \frac{p^2}{2} + \frac{a}{\epsilon}\frac{q^2}{2}$ and 
the corresponding unitary operator $U_\epsilon = \text{e}^{-i\epsilon [\frac{p^2}{2} + \frac{a}{\epsilon}\frac{q^2}{2}]}= \text{e}^{-i[\epsilon \frac{p^2}{2} + a\frac{q^2}{2}]} \rightarrow  \text{e}^{-ia\frac{q^2}{2}}$ for $\epsilon \rightarrow 0$. In agreement with the Baker formula \cite{BCH70}:   
 
\begin{equation}
\label{Baker}
{e}^{\lambda \text{\footnotesize A}} B \text{e}^{-\lambda \text{\footnotesize A}} = B + \lambda [A,B]
+\frac{\lambda^2}{2!}[A,[A,B]]+\frac{\lambda^3}{3!}[A,[A,[ A,B]]]+\dots, 
\end{equation}
both operations lead to extremely simple transformations of the canonical pair $q, p$. The free evolution incidents generate: 

\begin{equation}
\label{ev1}
\text{e}^{i \tau \frac{p^2}{2}} \begin{pmatrix} q \\ p \end{pmatrix} \text{e}^{-i \tau \frac{p^2}{2}} = \begin{pmatrix} q + \tau p \\ p \end{pmatrix} = \begin{pmatrix}1 & \tau \\ 0 & 1 \end{pmatrix} \begin{pmatrix} q \\ p \end{pmatrix}
\end{equation}
while the potential shocks:

\begin{equation}
\label{ev2}
\text{e}^{i a \frac{q^2}{2}} \begin{pmatrix} q \\ p \end{pmatrix} \text{e}^{-i a \frac{q^2}{2}} = \begin{pmatrix} q \\ p-aq \end{pmatrix} = \begin{pmatrix}1 & 0 \\ -a & 1 \end{pmatrix} \begin{pmatrix} q \\ p \end{pmatrix}
\end{equation}

Within this scheme, an interesting operation is performed by  a pair of free evolution steps separated by an oscillator pulse: 

\begin{equation}
\label{ev3} 
\text{e}^{-i\tau \frac{p^2}{2}} \text{e}^{-i\frac{1}{\tau} \frac{q^2}{2}}\text{e}^{-i\tau \frac{p^2}{2}} \equiv F_\tau
\end{equation} 
It generates:
\begin{equation}
\label{ev4}
\begin{split}
q \rightarrow\tau p \\
p \rightarrow -\frac{1}{\tau} q
\end{split}
\end{equation}
which might be called the {\em squeezed Fourier transformation}. Curiously, an equivalent operation is performed by:

\begin{equation}
\label{ev5}
 \text{e}^{-i\frac{1}{\tau} \frac{q^2}{2}}\text{e}^{-i\tau \frac{p^2}{2}} \text{e}^{-i\frac{1}{\tau} \frac{q^2}{2}} \equiv F_\tau
\end{equation}
Henceforth the following product of the 6 unitary 
operations yields the transformation $q\rightarrow -q$ and $p \rightarrow -p$   (the parity operator) 

\begin{equation}
\label{Parity}
\text{e}^{-i\tau \frac{p^2}{2}} \text{e}^{-i\frac{1}{\tau} \frac{q^2}{2}}  \dots \text{e}^{-i\tau \frac{p^2}{2}}  \text{e}^{-i\frac{1}{\tau} \frac{q^2}{2}}  \equiv P, 
\end{equation}
whereas  the sequence of 12 unitary terms produces an evolution loop: 
\begin{equation}
\begin{split}
\label{loop1}
\underbrace{\text{e}^{-i\tau \frac{p^2}{2}} \text{e}^{-i\frac{1}{\tau} \frac{q^2}{2}} \dots \text{e}^{-i\tau \frac{p^2}{2}} \text{e}^{-i\frac{1}{\tau} \frac{q^2}{2}}}_{\text{12 terms}} \equiv 1
\end{split}
\end{equation}

An intriguing property of (\ref{loop1}) is that all 6 free evolution  exponents  arise with the same signs and so do the exponents of the  kick operations (a kind of non-perturbative Baker-Campbell-Hausdorff effect \cite{BCH70}). More  important aspect of (\ref{loop1}) is that it contains the free evolution intervals (see Fig.2). It means that the remaining eleven unitary operations must cause the free evolution inversion:

\begin{equation}
\begin{split}
\label{inve}
\underbrace{\text{e}^{-i\frac{1}{\tau} \frac{q^2}{2}} \text{e}^{-i\tau \frac{p^2}{2}} \dots\text{e}^{-i \tau \frac{p^2}{2}} \text{e}^{-i\frac{1}{\tau} \frac{q^2}{2}} }_{\text{11 terms}} \equiv \text{e}^{+i\tau \frac{p^2}{2}},
\end{split}
\end{equation}

\begin{figure}[H]
\centering
\includegraphics[width=3.5cm]{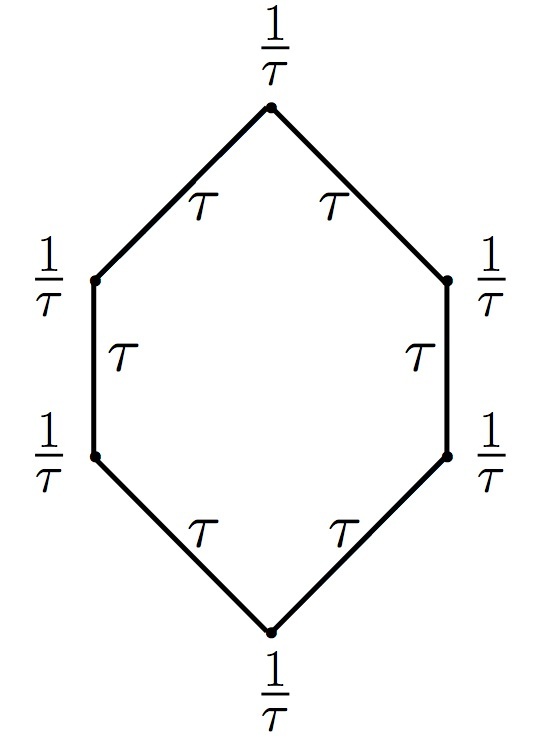}
\caption{The evolution loop formed by  12 elementary evolution operators. The  $\delta$-pulses of the attractive oscillator potential of amplitudes $1/\tau$ are represented by the hexagon vertices's, while the 6 sides symbolize the $\tau$-intervals of the free evolution. Each 3 consecutive operators yield the squeezed Fourier operation (\ref{ev3}-\ref{ev5}). Each 11 operations (6 consecutive oscillator kicks separated by 5 free evolution intervals) invert the free evolution.}
\label{eloop}
\end{figure}


If the idealized pulses could be indeed applied, the effect would be generated for every wave packet independently of its initial shape \cite{BM77}. Notice that in this way,  the loop mechanism (\ref{loop1}, \ref{inve})  predicted a part of the 1990 hypothesis about the 
quantum {\em time machine} \cite{Aha1990}.

The similar effects can be caused by elastic pulses with alternating signs. Their basic  element might be the sequence of 4 operators $S =\text{e}^{-i\tau \frac{p^2}{2}} \text{e}^{-ia\frac{q^2}{2}}\text{e}^{-i\tau \frac{p^2}{2}} \text{e}^{ia \frac{q^2}{2}} $, represented by the exponential functions of the nilpotent matrices  $ Q= \begin{pmatrix} 0 & 0 \\ 1 & 0 \end{pmatrix}$, $Q^\dagger = \begin{pmatrix} 0 & 1 \\ 0 & 0 \end{pmatrix}$:  
\begin{multline}
\label{4xS}
s = \text{e}^{\tau Q^\dagger}  \text{e}^{-aQ}  \text{e}^{\tau Q^\dagger}  \text{e}^{aQ} \\
=\begin{pmatrix}1 & \tau \\ 0 & 1 \end{pmatrix} \begin{pmatrix}1 & 0 \\ -a & 1 \end{pmatrix}  \begin{pmatrix}1 & \tau \\ 0 & 1 \end{pmatrix} \begin{pmatrix}1 & 0 \\ a & 1 \end{pmatrix} = \begin{pmatrix}1+ \tau a - \tau^2 a^2 & 2\tau-\tau^2 a \\ -\tau a^2 & 1-\tau a \end{pmatrix}
\end{multline}
The algebraic properties of $S$ depend again on the $\text{Tr}s=2- \tau^2 a^2$. If   
$ \tau^2 a^2>4\Rightarrow \text{Tr}s < -2$,  then the matrix  $s$ is of the type ({\bf III}) 
and the multiple pulse repetitions generate an unstable motion. However, if $0\neq \tau^2 a^2<4\Rightarrow -2<\text{Tr}s<2$ the matrix $s$ is of type ({\bf I}) and the repeated pulses yield a confined motion, including the possibility of the evolution loops. One of simplest cases occurs for  $ \tau^2 a^2=2\Rightarrow \text{Tr}s= 0 $; the Hamilton Cayley eq. for $s$ then implies $s^2=-\mathbb{1}\Rightarrow S^2\equiv -1$   
and so,  the pulse pattern $S$, if repeated,  creates the 16-step evolution loop,  in which the sum of the  oscillator pulses cancels, but the effects don't: 

\begin{equation}
\label{loop2}
[ \text{e}^{-i \frac{\sqrt{2}}{\tau}\frac{q^2}{2}} \text{e}^{-i \tau \frac{p^2}{2}} \text{e}^{+i \frac{\sqrt{2}}{\tau}\frac{q^2}{2}} \text{e}^{-i \tau \frac{p^2}{2}}]^4 \equiv 1
\end{equation} 

\bigskip

An elementary algebra  permits as well to predict some more general effects such as the 'time squeezing', i.e.,  accelerating or slowing down the free evolution \cite{Loops, Control}. Moreover, some simple, asymmetric sequences of the oscillator pulses can produce the squeezing and/or magnification of  canonical variables. The most elementary such effects are achieved by two different 'squeezed Fourier' operations, $F_\alpha$ and $F_\beta$:
\begin{equation}
\label{squeez}
F_\alpha F_\beta \rightarrow \begin{pmatrix} 0 & \alpha \\ -\frac{1}{\alpha} & 0 \end{pmatrix} \begin{pmatrix} 0 & \beta \\ -\frac{1}{\beta} & 0 \end{pmatrix} = \begin{pmatrix} -\frac{\alpha}{\beta} & 0 \\ 0 & -\frac{\beta}{\alpha} \end{pmatrix}  =  \begin{pmatrix} -\sigma & 0 \\ 0 & -\frac{1}{\sigma} \end{pmatrix} 
\end{equation} 
where $\sigma = \frac{\alpha}{\beta}$. Some of these phenomena were independently observed in \cite{Ni, Royer}. An ample collection of more general squeezing operations was described by  Dodonov \cite{Dodo}; the manipulation of complex Hamiltonians see \cite{Oscar13}. The idea of  controlling the  finite dimensional qubit systems by deforming  the closed dynamical processes reappears also in the recent development \cite{Viola,Emma,Knill}. This does not yet exhaust  all interesting effects.  

\bigskip

{\em An 'attractive repulsion'.}

It turns out that some special forms of trapped motion can be created around the centers of repulsive potentials. Assume first of all, that the particle obeying \eqref{n1} is submitted to a sequence of $\delta$-pulses with amplitudes $\pm a$, separated by the identical $\tau$-intervals of free evolution, such that $\tau^2 a^2>4$. The Floquet matrix $s=u(2\tau)$  then has $\text{Tr}s<-2$ and the pulses are repellent. Suppose, however, that the same  pulse pattern coexists with a constant repulsive potential  $V_\kappa(x)=-\kappa^2  \frac{x^2}{2}$, $\kappa>0 $ \cite{WolfMex,DavDav}

{\bf Proposition 2.} For adequate $a, \kappa, \tau$, the repulsive potential  $V_\kappa(x)$, in presence of the repelling pulse pattern,  can trap the particle. 

{\bf Proof}. 
The motion of the system can be again expressed by evolution matrices.  The evolution generated by the repulsive $V_\kappa(x)$  in any $\tau$-interval between two subsequent  $\delta$-pulses is described by the exponential matrix:   

\begin{equation}
\label{repex}
\text{e}^{\tau \Lambda} =  \frac{\text{e}^{\tau \Lambda}+\text{e}^{-\tau \Lambda}}{2}+\frac{\text{e}^{\tau \Lambda}-\text{e}^{-\tau \Lambda}}{2\Lambda} \Lambda
\end{equation} 
Here $\Lambda =\begin{pmatrix} 0 & 1 \\ {\kappa}^2 & 0 \end{pmatrix}$ fulfills $\Lambda^2=\kappa^2 \mathbb{1}$; hence, all even powers of $\Lambda$ in \eqref{repex} can be replaced by the corresponding powers of $\kappa$, leading to the well known hyperbolic matrix: 

\begin{equation}
\label{repelmatr}
\text{e}^{\tau \Lambda} =\cosh\kappa \tau \mathbb1 + \frac{\sinh\kappa\tau}{\kappa}\Lambda = \begin{pmatrix} \cosh\kappa \tau &    \frac{\sinh\kappa\tau}{\kappa}  \\  \kappa \sinh\kappa\tau     & \cosh\kappa \tau \end{pmatrix} 
\end{equation} 
Suppose now, \eqref{repelmatr} is intertwinned with  the matrices  \eqref{ev2} of the elastic $\delta$-kicks equivalent to  $\text{e}^{\pm aQ}$. By superposing  the constant  
 repulsion   acting in $[0,2\tau]$ with two opposite  kicks $\pm a$, one thus obtains:  
\begin{multline}
\label{repulsion2}
u _\kappa(2\tau) =\text{e}^{\tau \Lambda} \text{e}^{-aQ} \text{e}^{\tau \Lambda}\text{e}^{aQ}=\\ \begin{pmatrix} (1-\frac{a^2}{2\kappa^2})\cosh2\kappa\tau+\frac{a^2}{2\kappa^2}-\frac{a}{2\kappa}\sinh2\kappa\tau   &  \frac{1}{\kappa}\sinh2\kappa\tau +\frac{a}{2\kappa^2}(\cosh2\kappa\tau-1) \\ (\kappa-\frac{a^2}{2\kappa})\sinh2\kappa\tau -\frac{a}{2}(\cosh2\kappa\tau-1)  & \cosh2\kappa\tau+\frac{a}{2\kappa}\sinh2\kappa\tau   \end{pmatrix}  
\end{multline}

\begin{figure}[H]
\centering
\includegraphics[width=6cm]{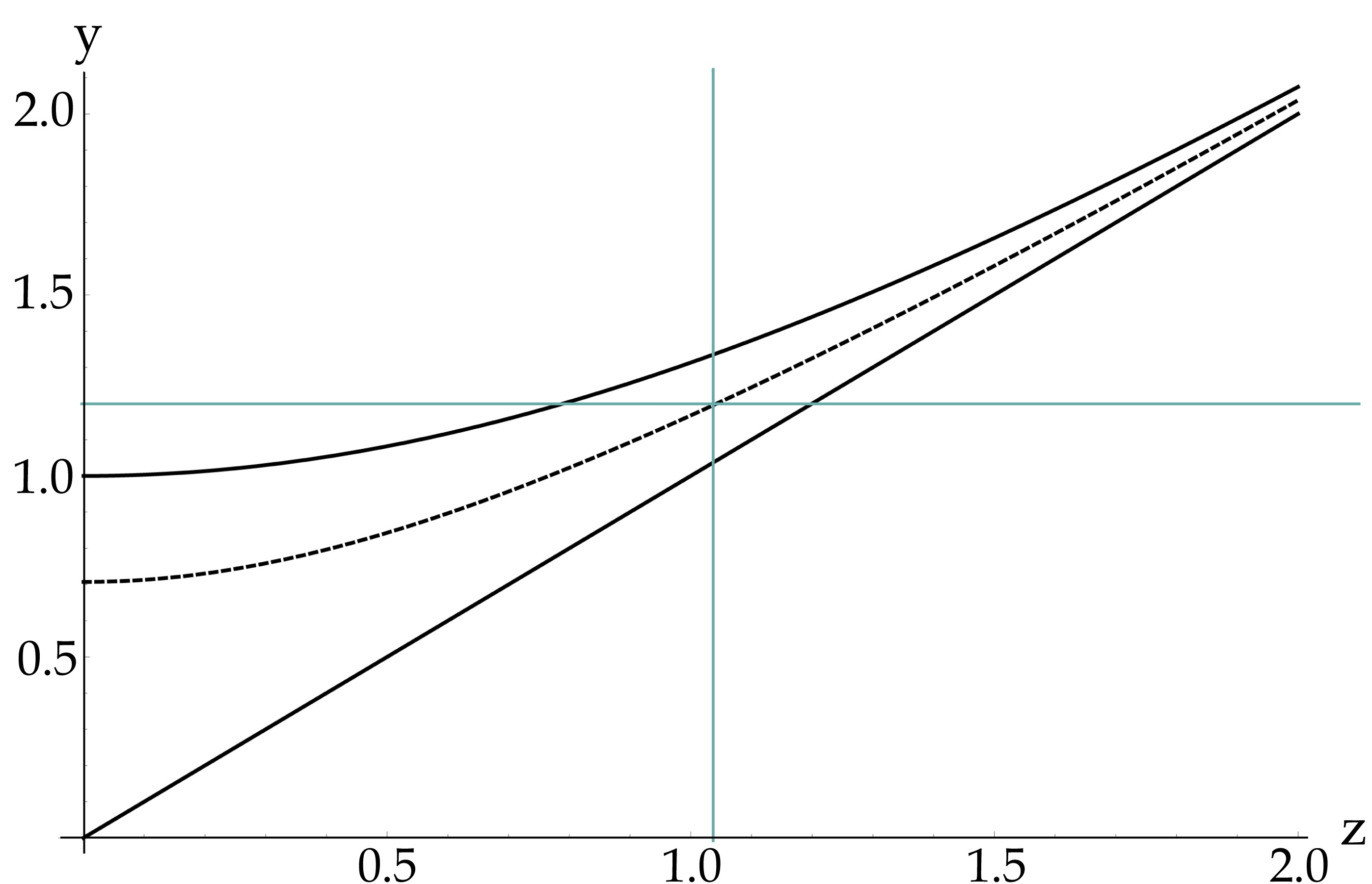}
\caption{The effect of the repulsive potential $V(x)=-\kappa \frac{x^2}{2}$ in a repelling sequence of  
the  elastic kicks   $\pm a$ at the time moments $n\tau, n=1,2,...$, on the parameter map $y=\frac{1}{2}a\tau$, $z=\kappa \tau $. The lower and upper borderlines mark the stability area where the evolution is confined. The dotted curve between marks the zeroes of \eqref{reptrace} where the  motions form octagonal loops. For $y=1.2$ the repulsive potentials with $z<0.79$ are still too weak, but for  $0.79<z<1.2$ they trap the motion. For  $z\simeq1.04296$ they generate  an evolution loop. } \label{fig9}
\end{figure}

with: 
\begin{equation}
\label{reptrace}
\operatorname{Tr}u_\kappa(2\tau) =2+(4-\frac{a^2}{\kappa^2})\sinh^2\kappa \tau
\end{equation} 

\bigskip

For $\kappa\rightarrow 0$ this reduces to $\text{Tr}u_0 (2\tau)=2-a^2\tau^2$, so  if the 'auxiliary variable' $y^2=\frac{1}{4}a^2\tau^2>1$, then  $u_0 (2\tau)$ is of type ({\bf III}) and the particle escapes. Yet, if the motion is assisted by an additional  repulsive potential where $z=\kappa\tau$ (the second auxiliary variable) satisfies $z^2<y^2<z^2(1+\frac{1}{\sinh^2z})$
 then the particle remains trapped (see Fig.3). $\square$
\bigskip

In particular, when   
\begin{equation}
\label{reploop}
z^2(1+\frac{1}{2\text{sinh}^2z}) =y^2
\end{equation} 

\bigskip

then the trace \eqref{reptrace} vanishes, the matrix $u _\kappa(2\tau)$  satisfies $u _\kappa(2\tau)^2=-1\Rightarrow u _\kappa(2\tau)^4=1$, and the particle is trapped in an  octagonal evolution loop represented on Fig.4, where the shadowy lines represent the elastic repulsion.

\begin{figure}[H]
\begin{center}
	\includegraphics[width=.40\textwidth]{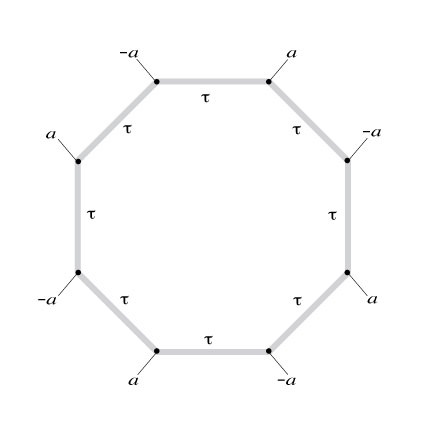}
\end{center}
\caption{The scheme of an atypical evolution loop in 1D, created by a sequence of 8 alternating oscillator 
pulses $\pm a$   in presence of a constant repulsive potential $V(x)=-\kappa \frac{x^2}{2}$, represented by the shadowy lines,  with $\kappa, \tau$ and $a$ satisfying \eqref{reploop}. In quantum case, the bound state is created around the repulsive center. Should the repulsive force disappear, the particle will escape. }  
\end{figure}

Even if one objects the S/F story about the elastic $\delta$-pulses, the evolution illustrated by Fig.4 is indeed an elementary equivalent  of a phenomenon known  for particles in Paul's traps moving  under the joint influence of two-component oscillator potentials,  $\beta (t)= \beta_0+\beta_1\text{sin}\omega t $. The stability areas on the 
($\beta _0, \beta_1$)-map are then determined by the  {\em Strutt diagram} \cite{Bender}. Each stability arm  extends toward the negative values of $\beta _0$ (repulsive oscillators) which may be intuitively expected to accelerate the  expulsion, but instead,  they protect the particle against the repelling action of the oscillating field. 

The effect shows also some similarity to the behavior of a charged particle moving in crossed electric  and magnetic fields in presence of a repellent obstacle in form of a rigid disk \cite{Repulsive}. In this scheme, the particle 'returns obsessively' to the obstacle: but should the repellent disk disappear,  it will  escape. The question, whether an analogous effect can exist for other,  
non-quadratic potentials is open. Should it occur for the repulsive Coulomb fields  
submerged in pulsating perturbations, it might contribute to a better understanding of some solid state effects  such as the mechanism of Cooper pairs. 
 
\bigskip

\section{Optical Equivalents: the true prehistory?}

While the realistic cases of  the idealized operations are still under discussion, it  was recently noticed that almost all quadratic control algorithms admit  a distinct interpretation. As reported by Wolf \cite{Historical,Wolf88}, the evolution  operations generated by the Hamiltonians (\ref{n1}) have some simple equivalents in  the  optical experiments on 1D optical bench. In particular, 
the $2\times2$  matrix  transformations (\ref{ev1}, \ref{ev2}, \ref{squeez}) of the canonical variables $q$, $p$ correspond exactly to the application of some typical optical instruments (microscopes, telescopes etc) well known in the geometrical optics. In fact, supposing that $q$ is the distance from the bench axis $z$ and $p=c \text{sin}\theta$, where $\theta$  is an angle between the light ray and the $z$ axis,  the matrix  (\ref{ev1}) describes  the optical images formed by the congruence of light rays propagating along the bench, where  $\tau$ is now the $z$-distance between the source and the image, while (\ref{ev2}) describes the action of a thin optical lens placed on the bench, the amplitude $a$ meaning the (positive or negative) lens curvature \cite{OpWolf}).

Both interpretations have certain gaps and advantages.  Thus, what in the dynamical language is an {\em evolution loop} (convenient for the dynamical manipulations), in optical terms is the simple reproduction of an optical image. The squeezing/amplification mechanisms produced by two or more oscillator shocks in quantum mechanics are easily interpretable as the applications of the microscope (or telescope) in  geometrical optics. (So, in a sense, the effects of the oscillator kicks were known already to Galileo).  Moreover, what was not so easy to predict in the dynamical language, i.e., the asymmetry of $\beta (t)$ needed to produce the squeezing, is immediately obvious at the optical level. (In fact, the symmetric apparatus could not  produce amplified or reduced images). Both, dynamical and optical techniques have also their specific imperfections.  In the terrain of optics, the applications of too close lenses with too big $\vert a \vert$  would mean that a part of one lens (if not the whole) must overlap with the interior of the other. More exact equivalents of the elastic pulses \eqref{ev2} were subsequently considered in works on optical signals in  dispersive fibers \cite{Agarwal} (more recent studies, c.f. also \cite{Rocio}). In quantum mechanics of particles driven  by potentials in ion traps the exact application of $\delta$ kicks is practicably impossible. Forgetting even about the finite resistance of the  trap walls, no infinite energy shocks can be truly engineered.  Yet, despite all their imperfections, the idealized optical (or dynamical) operations might be of interest for some unfinished fundamental discussions.  

\section{The Time-Energy uncertainty?}

\bigskip

Indeed, it is enough to remind the doubts concerning the "time-energy uncertainty principle". 
It looks that the  "squeezed Fourier" transformations can bring some new elements  
to the list of critical arguments. The first objections against the (too vebal) interpretations  of Landau 
and Peierls \cite{Landau} supporting this principle,  appeared in the study of  
Aharonov and Bohm in 1961 \cite{Aha61}. The doubts  
returned in \cite{Aha2000} and other papers. Further  arguments against too  
dogmatic formulations were collected in  Aharonov, Massar and Popescu 
who argue that an arbitrarily exact measurement of the energy of a quantum system 
can be performed in an arbitrarily short time, provided that "the measurement is brutal". Their illustration was the spin measurement \cite{Aha2002}.  As it seems, the  pulse patterns of Sec.4 bring the next illustration to the same idea. 

\bigskip

  Indeed, suppose that the two $\delta$-pulses divided only by a very short time interval 
$\tau$, or alternatively, one $\delta$-pulse of the oscillator potential 
between  two infinitesimal  intervals of the free evolution, perform the "squeezed Fourier 
transformation" \eqref{ev3} of a free particle propagating initially with an energy 
$\frac{p^2}{2m}$. After the operation, the unknown (classical or quantum) momentum $p$ is converted into the new particle position $\tilde{q}=\tau p$. Could such transformation be produced, it would 
be no longer necessary to detect 
directly the particle energy, e.g., by observing its collision with a heavier microobject 
 \cite{Landau}. It would be enough to measure its new position $\tilde{q}$. Whatever the technical difficulties, there is no fundamental law which would forbid to determine the new position $\tilde{q}$ in an arbitrarily short time.\footnote{What might awake some doubts is the fact that in such measurement the particle 
position $\tilde{q}$ would be used to detect 
its momentum $p$. Is it not in conflict with the position-momentum uncertainty? Yet, 
it is not, since  $\tilde{q}$ and $p$ are the particle position and momentum in different time moments. In the orthodox quantum oscillator, the position  measurement at some time moment $t$ is equivalent to the momentum measurement at $t-\frac{T}{4}$, where {\footnotesize T} is the oscillator period (compare with  the 'nondemolishing measurement' \cite{Thorne}). } 

\bigskip

What one can still object,  is that our particular prescriptions of generating \eqref{ev3} offer an inconvenient relation between the finally measured $\tilde{q}$ and the momentum $p$ prior to the applied operation. The $\tilde{q}=\tau p$ implies $\Delta \tilde{q}=\tau \Delta p$, so if the operation time $\tau$ is very short, then little errors in measuring $\tilde{q}$ will correspond to much greater errors in $p$. It looks almost as the vengeance of the time-energy uncertainty. Yet, it isn't! In fact, $\tau$  is a $c$-number parameter defining the time of an external operation, valid in classical as well as in quantum theory, an authentic {\em external time} in sense of Aharonov and Bohm \cite{Aha61}, and $\Delta \tilde{q}$ {\em is not} limited by  any universal constant. Moreover, still accepting the pulse solutions, the dynamics of the time dependent Hamiltonians (\ref{n1}) can   offer much better measurement methods. Following the optical analogy \cite{Historical}, we could call it: 
\\~\\
{\em The "Fourier microscope"}. 

Indeed, it is enough to apply three consecutive squeezed Fourier transformations, to obtain a unitary operator with a more convenient transformation matrix: 

\begin{equation}
\label{microscope}
F_\mu F_\gamma  F_\mu \rightarrow \begin{pmatrix} 0 & \mu \\ -\frac{1}{\mu} & 0 \end{pmatrix} \begin{pmatrix} 0 & \gamma\\ -\frac{1}{\gamma} & 0 \end{pmatrix}  \begin{pmatrix} 0 & \mu \\ -\frac{1}{\mu} & 0 \end{pmatrix} = \begin{pmatrix} 0 & b  \\ -\frac{1}{b} & 0 \end{pmatrix}
\end{equation} 

where the new coefficient $b =- \frac{\mu^2}{\gamma}$ is unlimited. Thus, taking $\mu = 1$, one would obtain a "squeezing mechanism" where the final position $\tilde{q}= b p$ for large $b$ leads to insignificant errors in $\Delta p= \frac{1}{|b|}\Delta \tilde {q}$  even if the $\tilde{q}$ measurement is far from perfect. The problem of an efficient empirical design is still open.  One of the simplest ways of measuring the microparticle position is to let it interact with a lattice of mesoscopic absorption centers (e.g., grains of   photographic emulsion), one of which turns dark, marking the new particle position $\tilde{q}$. As the particle is projected into a lattice, the $\Delta \tilde{q}$ errors will be of order of magnitude of the lattice distance $\Delta l$ between the neighboring mesoscopic centers, but  anyhow, the initial particle momentum $p$ can be determined with an arbitrarily  small error   $\Delta p = \gamma \Delta l$ if $\gamma$ is little enough. While technical difficulties still exists, they do not seem a fundamental obstacle. 
\bigskip

The operations considered above are singular and concern the states in 1 space dimension. Their generalization in 2D and 3D, employing as well  $\delta$-kicks, were considered by  Fernandez \cite{David92}, including 
the comments on  the {\em time-energy} uncertainty, though the results somehow escaped 
attention in  the noise of the research markets. Yet, all effects  described above are just  idealized forms of some natural phenomena which 
might be of technical interest. In particular, the merits of the evolution loops were noticed again  in the control problem of finite dimensional spaces (qubits) \cite{Harel},  recognized more generally in \cite{Emma}. The continuous evolution affected by sequences of sharp  $\delta$-kicks, under the name of {\em decouplers}, are now studied for the spin systems,  as one of promising quantum control methods \cite{Harel, Viola} (the so called {\em 'bam-bam}-control'), though less violent methods are also considered \cite{Knill, Pasini}. In the infinite dimensional $L^2(\mathbb{R})$ the  soft alternatives of the potential kicks can be no less interesting.  

\section{The elementary algebraic solutions}

As the matter of fact, the techniques of applying the  non-singular, bounded fields 
 (e.g. in form of rectangular steps)  allowed already to design an ample family of dynamical 
operations including the squeezing, distorted free evolution, etc. \cite{Loops,Control,Dav11,Dallas}). 
The possibility of approaching the same effects by softly varying, differentiable  fields (without any steps)  was considered in \cite{Gerardo,Control}, then elaborated in computer studies for charged particles submerged  in harmonic pulses \cite{PhDel,FDel,Ale}.  As subsequently found, the control operations are significantly simplified  if the field amplitude $\beta (t)$ is symmetric with respect to the center of the operation interval   \cite{Pasini, Ale1,Ale2}. We shall show now, that in the symmetric generation mechanisms the matrices (\ref{n2}-\ref{n8}), together with the corresponding  'driving amplitude' $\beta(t)$ can be expressed exactly (without perturbations!) in terms of a single real function which, appart of details, may be fixed at will. Without pretending to be a major mathematical discovery, it facilitates significantly the task of programing the  dynamical operations.  Indeed, one has:  
\bigskip

{\bf Proposition 3.} Consider a nontrivial operation interval $[-\text{\footnotesize{T}},\text{\footnotesize{T}}]$ with the quadratic potential $\beta(t) \frac{q^2}{2}$ and suppose, $\beta(t)$ is  bounded, piecewise continuous and symmetric $\beta(t) = \beta(-t)$. Whenever $u= u(t,-t)$, for $t\in [0, \text {\footnotesize{T}}]$, reaches the stability threshold with $\text{Tr}u=\pm 2$, then either $ u$ or -$u$ adopts one of the forms \eqref{ev1} or \eqref{ev2}, imitating the results of a simple or distorted free evolution, or else, of a sharp oscillator kick. Moreover,  the  
evolution matrix $u(t,-t)$ and the field amplitude $\beta(t)$ in the expanding family of intervals $[-t,t]$ (where  $t\leq\text{\footnotesize{T}}$)  can be written explicitly in terms of $\theta (t)= u_{12}(t,-t)$, which may be choosen at will everywhere except its zero points. (In what follows, whenever there will be no reasonable doubt, we shall simplify the notation, writing just $u(t)$ instead $u(t,-t)$ and $u_{kl}(t)$ instead of $u_{kl}(t,-t)$). 

{\bf Proof.}
 Due to the symmetry of the Hamiltonians   $H(t) = H(-t)$, the   unitary evolution operators $U(t,-t)$ for the expanding intervals  $[-t,t]$ satisfy: 

\begin{equation}
\label{Sym1}
i \frac{d}{dt}U(t,-t) = H(t)U(t,-t)+U(t,-t)H(t)
\end{equation}
Hence, the corresponding evolution matrix $u(t) = u(t,-t)$ is differentiable and fulfills:
\begin{equation}
\label{Sym2}
\frac{du}{dt} = \Lambda(t)u + u\Lambda(t)
\end{equation} 
Since $\Lambda(t)$ is given by (\ref{n4}), this becomes
\begin{equation}
\label{Sym3}
\begin{split}
\frac{du}{dt}& = \begin{pmatrix} u_{21}-\beta u_{12} & \text{Tr} u \\ 
-\beta\text{Tr}u & u_{21}-\beta u_{12}
\end{pmatrix} \\
&=(u_{21}-\beta u_{12})\mathbb{1} + \text{Tr}u \begin{pmatrix} 0 & 1 \\ -\beta & 0\end{pmatrix} 
\end{split}
\end{equation}
Therefore,
\begin{equation}
\label{Sym4}
\frac{d}{dt} (u_{12}u_{21}) = \text{Tr}u (u_{21}-\beta u_{12}) = \text{Tr}u \frac{1}{2}\frac{d}{dt} \text{Tr}u = \frac{1}{4} \frac{d}{dt} (\text{Tr}u)^2
\end{equation}
and integrating:
\begin{equation}
\label{Sym5}
\frac{d}{dt}\left[ u_{12}u_{21}-\frac{1}{4} (\text{Tr} u)^2 \right] = 0 \Rightarrow u_{12}u_{21} - \frac{1}{4}(\text{Tr}u)^2 = C = const.
\end{equation}
To determine $C$ it is enough to take $t= 0$. The initial condition $u(0,0) = \mathbb{1}$ then tells that $C=-1$, and so: 
\begin{equation}
\label{Sym6}
u_{12}u_{21} = \frac{1}{4}(\text{Tr}u)^2 - 1 \quad t \in [-\text{\footnotesize{T}},\text{\footnotesize{T}}]
\end{equation}

Hence, whenever the symmetric  evolution matrix $u(t)=u(t,-t)$ reaches 
the threshold values  Tr$u =\pm2$ 
(the case II of our classification), (\ref{Sym6}) implies that either $u_{1,2}$ or $u_{2,1}$ (or both) must vanish, leading to the canonical transformations (\ref{Sym3}) which  simulate the oscillator kicks, 
incidents of distorted free evolution, or just one of  the evolution loops (c.f.~\cite{Ale1, Ale2}), all of them with or without the simultaneous parity transformation.

These facts are already a significant  advantage of  (\ref{Sym3}) which, in  
 addition,  provides an elementary solution of the inverse evolution problem,  permitting to reconstruct the entire  evolution matrices $u(t,-t)$  in terms of one arbitrary function $\theta = u_{1,2}(t)$ which  determines simultaneously  the driving pulse $\beta(t)$  in the expanding intervals $[-t,t]$. 

Indeed, 
(\ref{Sym3}) implies $\frac{du_{11}}{dt} =\frac{du_{22}}{dt}= u_{21} - \beta u_{12}$ and since the initial condition at $t=0$ is $u_{11}(0) = u_{22}(0) =1$, then $u_{11}=u_{22}$ in all  intervals $[-t,t]$ ($t\leq\text{\footnotesize{T}}$). In view of (\ref{Sym3}) this means that $u_{11}(t) = u_{22}(t) = \frac{1}{2} \operatorname{Tr}u=\frac{1}{2} \theta'(t)$. In turn, since $u$ is simplectic, then $(\frac{1}{2}\theta')^2 - \theta u_{2,1}=1$ and the remaining matrix element $\alpha = u_{2,1}$  is determined as: 
  
\begin{equation}
\label{alpha}
\alpha = \frac{(\frac{1}{2}\theta')^2 - 1}{\theta}
\end{equation}

with the pulse shape $\beta(t)$  defined in terms of $\theta$ as well: 

\begin{equation}
\label{beta}
\beta = -\frac{\theta''}{2\theta}+ \frac{(\frac{1}{2}\theta')^2 - 1}{\theta^2}
\end{equation}

These expressions grant  that the matrix eqs. (\ref{Sym3}) are  fulfilled, yielding an 
exact solution of the symmetric 
evolution problem for the family of the expanding intervals $[-t,t]$. $\square$
\bigskip

 Unless stated otherwise the subsequent remarks concern the evolution matrices in the hypothetical symmetry interval. 
{\em Observation}. Since $\beta (t)$ in the reported algorithm is even, it is enough to postulate $\theta$ 
and find $\beta$ from \eqref{beta} for $t\geq0$, and then to reconstruct $\beta$ for $t<0$ by parity argument  
$\beta(-t)=\beta(t)$. 
\bigskip

The demand to obtain a physically interpretable result with nonsingular, bounded $\beta$-pulse permits one to use  an arbitrary twice differentiable $\theta$ with bounded $\theta''$, limited only by certain auxiliary conditions. In particular, it might be of interest to look for the pulse amplitude $\beta$ vanishing 
in some finite or infinite  intervals on the $t$-axis. If not vanishing everywhere,  $\beta$ cannot be analytic on $\mathbb{R}$; however, it can be switched on or off softly, to remain continuous together with several 
derivatives. The simplest cases of vanishing $\beta$ occur in sub-intervals where $\theta'(t)=\pm 2$, 
(i.e., when the graphic of $\theta$ on ($\theta, t$)-plane sticks to one of the straight lines {$\theta = \pm 2t$+const}). However, this is not the only case.  
\bigskip

{\bf Proposition 4.} If $\beta(t)=0$ in some interval $\mathcal{I} =[t_o,t_1]\subset \mathbb{R}_+$, then $\theta$ is quadratic in $\mathcal{I}$, i.e. $\theta(t)=\text{\footnotesize a}t^2+\text{\footnotesize b}t+\text{\footnotesize c}$, with $\text{\footnotesize b}^2-4\text{\footnotesize ac}=4$. 
 
{\bf Proof.} Note, that  $\theta$ can vanish only in isolated points, not in subintervals,  and not together with $\theta'$, otherwise 
the matrix $u$ could not be simplectic. Hence, \eqref{beta} implies: 
\begin{equation}
\label{begamma}
\beta \theta^2= -\frac{1}{2}\theta''\theta+ \frac{1}{4}\theta'^2 - 1 \\~~\\ \Rightarrow\\~~\\ (\beta \theta^2)'= -\frac{1}{2}\theta'''\theta 
\end{equation} 
and so 
\begin{equation}
\label{3grado}
 \beta' \theta + 2\beta\theta'= -\frac{1}{2}\theta''' 
\end{equation} 
 
As a consequence, in any interval where $\beta=0$ 
there is $\theta'''=0 \Rightarrow \theta (t)=\text{\footnotesize a}t^2+\text{\footnotesize b}t+\text{\footnotesize c}$. Moreover, by entering into \eqref{begamma} one sees that  $\text{\footnotesize b}^2-4\text{\footnotesize ac}=4$. $\square$ 
 
\bigskip
Differently than $\beta$, the function $\theta$ cannot vanish in non-trivial intervals, but its vanishing at some particular points might be of interest.  
Below,  $t=0$ is the symmetry center of $\beta$. Obviously,  $u(0,0) = \mathbb{1}\Rightarrow$ $\theta(0) =0, \theta'(0) =2$ and 
$\alpha(0) = u_{2,1}(0)= 0 \Rightarrow \theta''(0) =0$. If moreover $\theta$ is three times differentiable at $t=0$, 
then its third derivative $\theta'''$ defines the value of the driving amplitude $\beta(0)= -\frac{1}{8}\theta'''(0)$ at the symmetry point (all this following from  \eqref{3grado},   Proposition 4).  

Apart of $t=0$,  $\theta (t)$  can have other null points in which the consistency conditions are 
more relaxed, corresponding to distinct dynamical effects. In consequence, the types of the operations 
defined by $\theta (t)$ in any symmetry interval $\text{\footnotesize{I}}=[-\text{\footnotesize{T}},\text{\footnotesize{T}}]$ can be deduced just from very simple data 
including $\theta$ and its derivatives. In particular, the 'distorted free evolution', the 'squeezed Fourier' and the simulation od a sudden $\delta$-kick of the oscillator potential obey the  
Propositions 5, 6, 7: 
\bigskip

{\bf Proposition 5.} If $\theta(\text{\footnotesize{T}})\neq 0$ and  $\theta'(\text{\footnotesize{T}} )= \pm 2$  then  the matrix  $u=u(\text{\footnotesize{T}},-\text{\footnotesize{T}})$ adopts one of the characteristic  shapes of  the {\em  deformed free evolution}, i.e.: 
\begin{equation} 
\label{time}
 \theta'(\text{\footnotesize{T}} )= 2 \Rightarrow 
u= \begin{pmatrix} 1 & \tau \\ 0 & 1 \end{pmatrix}, \\~~~\\  \theta'(\text{\footnotesize{T}} )= -2 \Rightarrow  u=\begin{pmatrix} -1 & \tau  \\ 0 & -1 \end{pmatrix}  
\end{equation} 
with the  'distorted time'  
$ \tau =\theta(\text{\footnotesize{T}} )$ or $ \tau =-\theta(\text{\footnotesize{T}} )$ respectively. Moreover, if  $\theta''(\text{\footnotesize{T}} )=0$, then the driving amplitude $\beta(t)$ vanishes softly,  $\beta(\text{\footnotesize{T}}) = 
\beta'(\text{\footnotesize{T}} )=0$, in the limits of the operation interval. 
  
{\bf Proof.} Indeed, the diagonal terms of $u$ in both cases are $u_{11}=u_{22}= \text{Tr}u =\pm 1$. 
Hence,  if $\theta (\text{\footnotesize{T}} )\neq 0$ the  nominator of \eqref{alpha} vanishes and so, 
$\alpha = u_{21}=0$ implying \eqref{time}. Note also that whenever the evolution yields the second matrix of \eqref{time}, then its repetition $u^2$ in $[-\text{\footnotesize{T}},3\text{\footnotesize{T}}]$ will give  $u^2 =\begin{pmatrix} 1 & -2 \tau  \\ 0 & 1 \end{pmatrix}$; so if $\tau$ in \eqref{time} was positive, 
then $u^2$ generates the free evolution inversion with the effective time $-2\tau$. $\square$
\bigskip

{\bf Proposition 6.}  If $\theta (\text{\footnotesize{T}} ) \neq 0$, $\theta' (\text{\footnotesize{T}} )=0$ and $\theta''(\text{\footnotesize{T}} )\theta(\text{\footnotesize{T}} ) = -2$, then the matrix $u$ produces the {\em squeezed Fourier transformation} \eqref{microscope}  
with $b=\theta(\text{\footnotesize{T}})$. If in addition $\theta'''(\text{\footnotesize{T}} )=0$, then 
the driving amplitude $\beta(t)$ vanishes softly (with the continuous first derivative) outside of the operation interval.  
 
{\bf Proof.} Indeed, if  $\theta' (\text{\footnotesize{T}} )=0$, then the corresponding matrix $u$ has zeros on the diagonal, with $u_{12}=b=\theta(\text{\footnotesize{T}})$. Moreover, if $\theta''(\text{\footnotesize{T}} )\theta(\text{\footnotesize{T}} ) = -2 \Rightarrow \theta''(\text{\footnotesize{T}})=-\frac{2}{b}$, then \eqref{begamma} implies that $\beta(\text{\footnotesize{T}})=0$, and if $\theta'''(\text{\footnotesize{T}})=0$, then also $\beta'(\text{\footnotesize{T}})=0$. $\square$
\bigskip

{\bf Proposition 7.}
 When $\theta (\text{\footnotesize{T}}  )=0$ for $\text{\footnotesize{T}}  \neq 0$, the nonsingularity of $\beta$ and $\alpha$ 
at $t=\text{\footnotesize{T}}  $ requires $\theta'(\text{\footnotesize{T}}  )= \pm2$. Then, $\beta(\text{\footnotesize{T}} ) = -\frac{1}{8}\text{sgn}[\theta'(\text{\footnotesize{T}}  )]\theta'''(\text{\footnotesize{T}} )$ and $\alpha(\text{\footnotesize{T}} )= \frac{1}{2}\theta''(\text{\footnotesize{T}}  )$. In particular, if $\theta'''(\text{\footnotesize{T}})=0$ and $\theta''(\text{\footnotesize{T}})=-2a\neq 0$ then $\beta (t)$ vanishes  at the extremes  $ \pm \text{\footnotesize{T}}$ and $u(\text{\footnotesize{T}})$ simulates the effect of the $\delta$-pulse $\beta(t)=a\delta(t)$. 

{\bf Proof.} Indeed, due to non-singularity and vanishing of $\theta $ at $t =\text{\footnotesize{T}} $, the eq.\eqref{begamma} 
implies $\frac{1}{4}\theta'^2-1=0\Rightarrow \theta'(\text{\footnotesize{T}} ) =\pm 2$. Hence, \eqref{3grado} reduces to $\pm 4\beta(\text{\footnotesize{T}} )= - \frac{1}{2}\theta'''(\text{\footnotesize{T}} )$. The $\alpha(\text{\footnotesize{T}} )$ in \eqref{alpha} is the nonsingular limit of a fraction in which both nominator and denominator tend to 0 for $t\rightarrow \text{\footnotesize{T}} $; the eq. \eqref{begamma} showing that  $\alpha=\frac{1}{2}\theta''(t)+\beta(t)\theta(t)\rightarrow \frac{1}{2}\theta''(\text{\footnotesize{T}} )=\alpha(\text{\footnotesize{T}})=-a$ for $t\rightarrow \text{\footnotesize{T}} $. $\square$

\bigskip
{\em The elementary models}. While the best choice of $\theta (t)$ from the point of view of the laboratory techniques is still  an open problem, one can notice the existence of very simple models showing how could  the dynamical operations \eqref{time} and \eqref{microscope} be achieved in either  fast or slow way. The choice of $\theta(t)=2\frac{\text{sin}\omega t}{\omega}$, of course,  brings no novelty, leading to  $\beta(t)=\omega^2= const$ of the traditional, time independent oscillator \eqref{n1}. However, already the simple polynomial modification: 

\begin{equation} 
\label{polymodel}
\theta (t) = 2t - \theta_3  t^3  - \theta_5  t^5 - \theta_7 t^7 - \theta _9 t^9
\end{equation}
 is sufficient  to design the examples of all operations of Propositions 5-7 in either short or long  intervals $[-\text{\footnotesize{T}},\text{\footnotesize{T}}]$. Thus, to generate the incidents of  'distorted free evolution', one needs:  
\begin{equation} 
\label{invefree}
\theta (\text{\footnotesize{T}}) = b, \\~~~\\ \theta' (\text{\footnotesize{T}}) =  \pm 2, \\~~~\\\theta'' (\text{\footnotesize{T}}) =0,\\~~~\\ \theta''' (\text{\footnotesize{T}}) = 0
\end{equation}
leading  to a 
systems of 4 linear,  equations for the 'new amplitudes' $\Theta_k = \theta_k \text{\footnotesize{T}}^k$, $k=3,5,7,9$, where the matrix $\text{{\bf A}}$ is the same for all operation intervals. The cases $ \theta' (\text{\footnotesize{T}}) =  \pm 2$ of \eqref{invefree} yield $\nu=0$ or  $\nu=4$ of the inverted or accelerated free evolution (see Fig.5.):

\begin{equation}
\label{matrix}
\text{{\bf A}}  \Theta=\text{{\bf A}}\begin{pmatrix} \Theta_3\\ \Theta_5 \\ \Theta_7 \\ \Theta_9 \end{pmatrix} =\begin{pmatrix} -b+2\text{\footnotesize T}\\ \nu \text{\footnotesize T}\\ 0 \\ 0 \end{pmatrix};  
\\~~~~\\ \text{{\bf A}} = \begin{pmatrix} 1 & 1 & 1 & 1 \\ 3 & 5 & 7 & 9\\3 & 10 & 21 & 36\\ 1 & 10 & 35 & 84  \end{pmatrix} 
\end{equation}

The operations of the 'distorted time' \eqref{time} could not occur for the constant oscillator potentials $\beta_0 \frac{q^2}{2}$, though can be generated by $\beta(t)$ in form of finite steps (see the maps in \cite{Loops,Control,Dav11}). They also occur in harmonic fields, though the results require a computer study \cite{Ale,Ale1,Ale2}. Here, they can be given by the exact formula \eqref{matrix}.
 In turn, the squeezed Fourier transformation \eqref{microscope} is generated by time independent 
oscillator potentials, but to obtain a large $b$ it needs a long time of waiting. Here, it requires only: 

\begin{equation} 
\label{squeeF}
\theta (\text{\footnotesize{T}}) = b, \\~~~\\ \theta' (\text{\footnotesize{T}}) =  0 \\~~~\\\theta'' (\text{\footnotesize{T}}) =-\frac{2}{b},\\~~~\\ \theta''' (\text{\footnotesize{T}}) = 0
\end{equation}
where $0\neq b \in \mathbb{R}$. Within the polynomial model \eqref{polymodel} it means:
\begin{equation} 
\label{polyF}
\text{{\bf A}}  \Theta=\text{{\bf A}}\begin{pmatrix} \Theta_3\\ \Theta_5 \\ \Theta_7 \\ \Theta_9 \end{pmatrix} =\begin{pmatrix} -b+2\text{\footnotesize T}\\ 0 \\ \frac{2\text{\footnotesize T}^2}{b}\\ 0 \end{pmatrix}
\end{equation}
leading to a family of $\theta(t)$'s illustrated on Fig.6.

\begin{figure}[H]
\begin{center}
	\includegraphics[width=.9\textwidth]{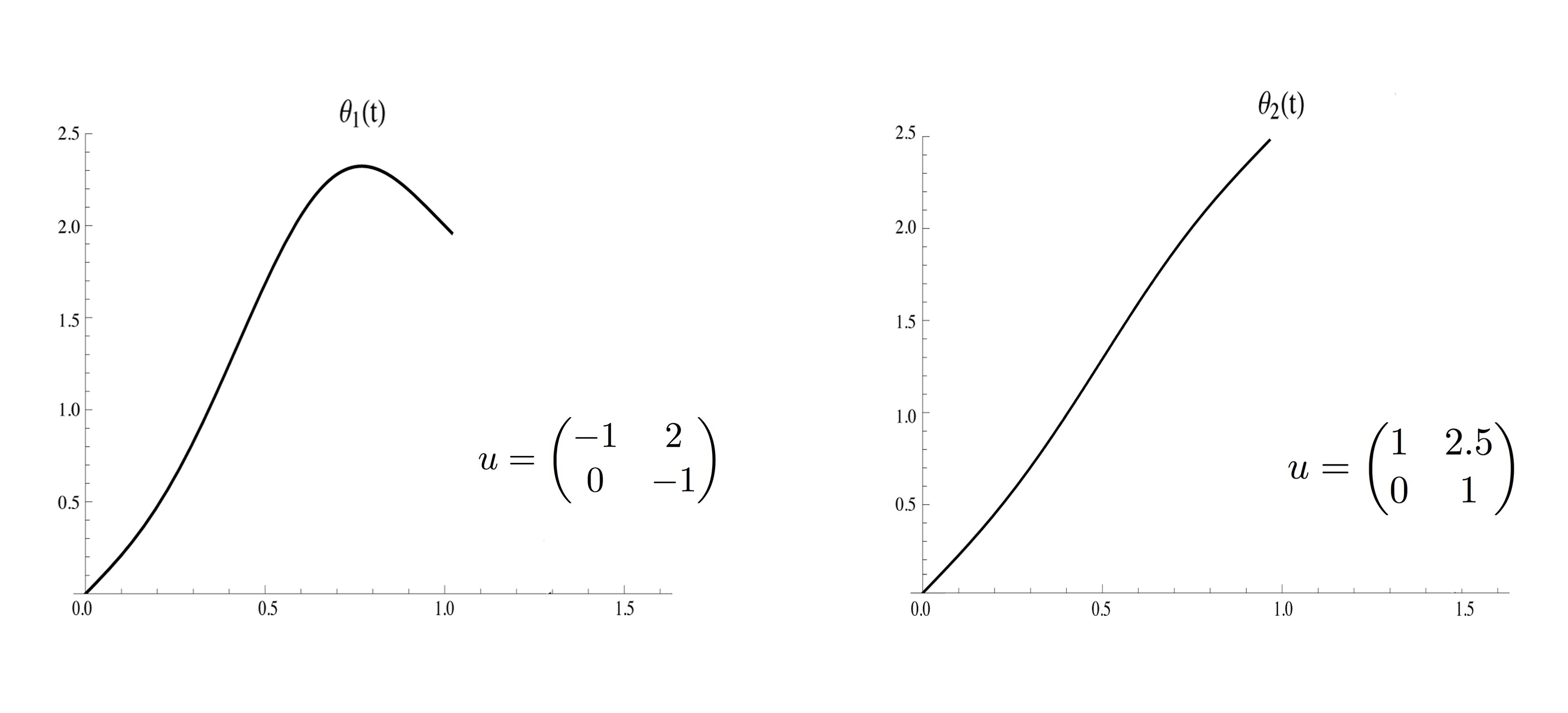}
\end{center}
\caption{The polynomial  $\theta_1(t) = 2t - \frac{41}{4} t^3 + \frac{93}{4} t^5 + \frac{71}{4} t^7 + \frac{19}{4} t^9$  with $\theta'_1(0)=2$, $\theta'_1(1)=-2$ yields parity$\times$free  evolution. If squared, it inverts the free evolution. In turn, $\theta_2= 2t-\frac{1}{32}\left(-105 t^3+189t^5-135t^7+35t^9\right)$ with  $\theta_2'(0)=\theta_2'(1)=2$ and $\tau=\theta_2(1)=2.5$  yields the free evolution acceleration} 
\label{fig:3}
\end{figure}

\begin{figure}[H]
\begin{center}
	\includegraphics[width=.50\textwidth]{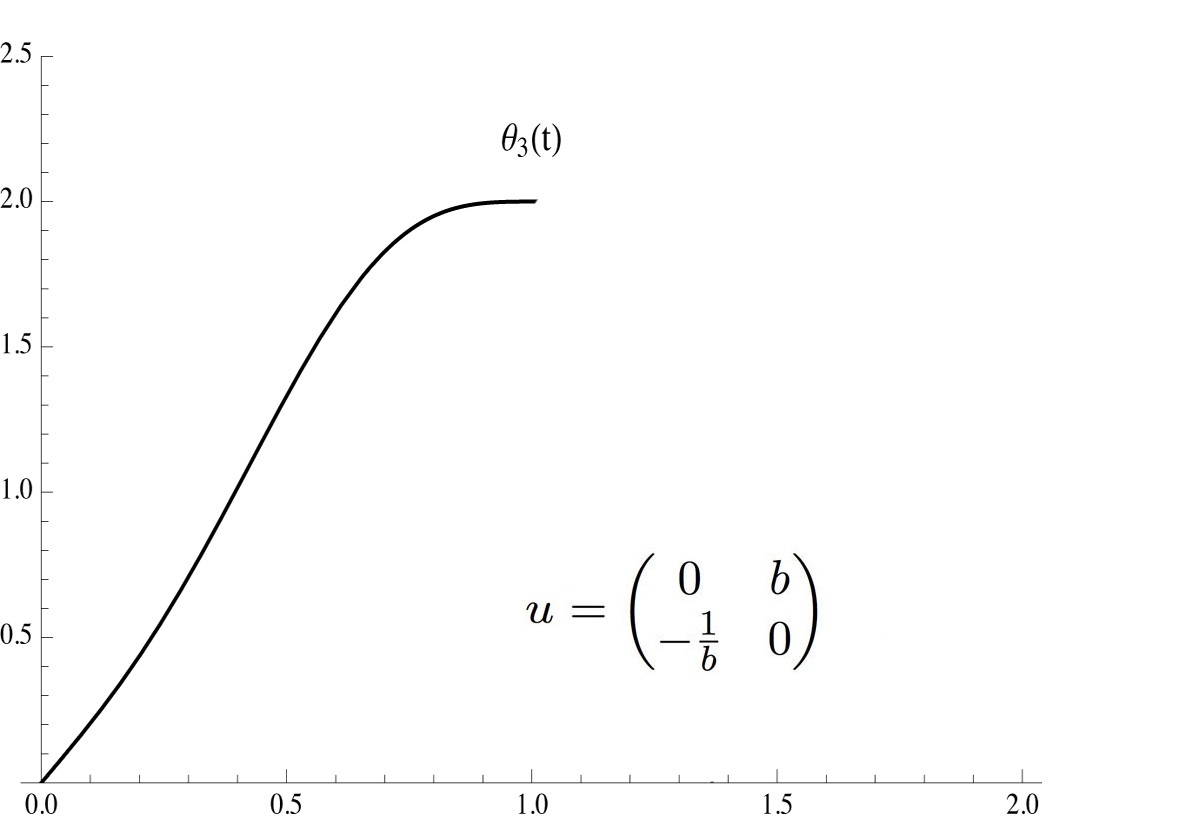}
\end{center}
\caption{One of typical shapes of $\theta_3(t) = 2t - \frac{3}{4} t^3 -2 t^5 - \frac{7}{4} t^7 + \frac{1}{2} t^9	$ permitting to generate the 'squeezed Fourier' in 
$[-1,1]$  for  $b=2$ and $\text{\footnotesize T} =1$ in \eqref{polyF}.}
\label{fig:3}
\end{figure}

The polynomials \eqref{polymodel} are, of course, not the unique models,    
but they permit to construct easily the sharp and soft alternatives of  the "time machine",  "Fourier microscope" etc.  The Fig.7  compares the shapes of the corresponding driving 
amplitudes $\beta (t)$. 

\begin{figure}[H]
\begin{center}
	\includegraphics[width=.50\textwidth]{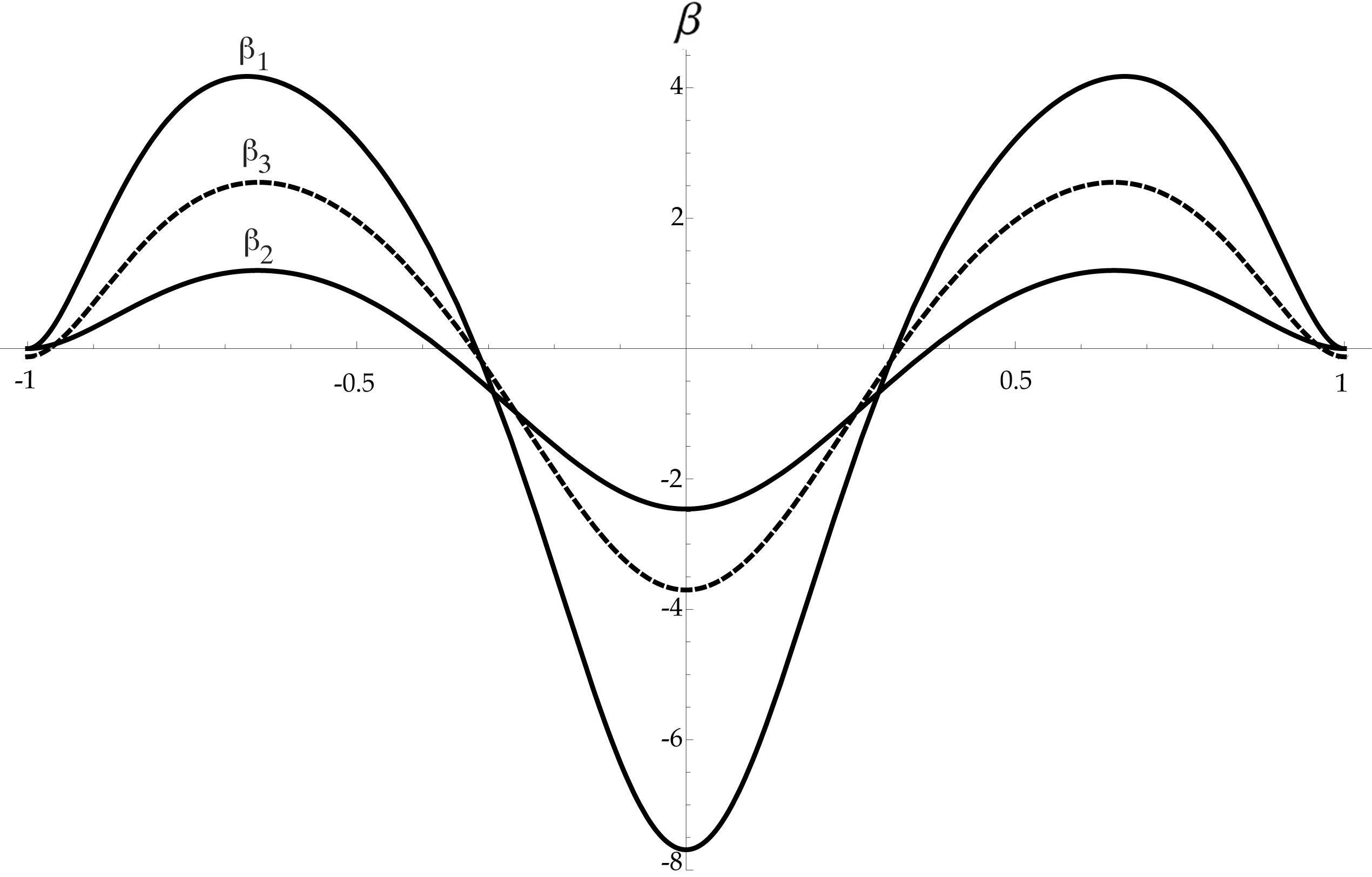}
\end{center}
\caption{The quantitative difference between two pulses $\beta_1$ and $\beta_2$ explains the 
qualitative difference between the inverted and accelerated  free evolution. In turn,  $\beta_3$ produces the squeezed Fourier operation represented on Fig.6.} 
\label{fig:3}
\end{figure}

\bigskip

Finally, the $\delta$-pulses $\beta(t)=a\delta(t)$, discussed in so many papers, admit now the soft polynomial  models \eqref{polymodel} with the boundary conditions obeying  the Proposition 7 (see Fig.8), i.e.:   

\begin{equation} 
\label{polyKick}
\text{{\bf A}}  \Theta=\begin{pmatrix} 2\text{\footnotesize T}\\ 4\text{\footnotesize T} \\ a\text{\footnotesize T}^2\\ 0 \end{pmatrix} \\~\\ \Rightarrow \\~~\\ \begin{pmatrix} \theta_3\\ \theta_5\\ \theta_7\\ \theta_9 \end{pmatrix}=\begin{pmatrix} \frac{23}{8\text{\footnotesize T}^2}+\frac{3a}{4\text{\footnotesize T}} \\ -(\frac{3}{8\text{\footnotesize T}^4}+\frac{2a}{\text{\footnotesize T}^3})  \\ -(\frac{7}{8\text{\footnotesize T}^6} - \frac{7a}{4\text{\footnotesize T}^5})  \\ \frac{3}{8\text{\footnotesize T}^8}-\frac{a}{2\text{\footnotesize T}^7}\end{pmatrix} 
\end{equation}
\bigskip

\begin{figure}[H]
\begin{center}
	\subfigure{\includegraphics[width=0.5\textwidth]{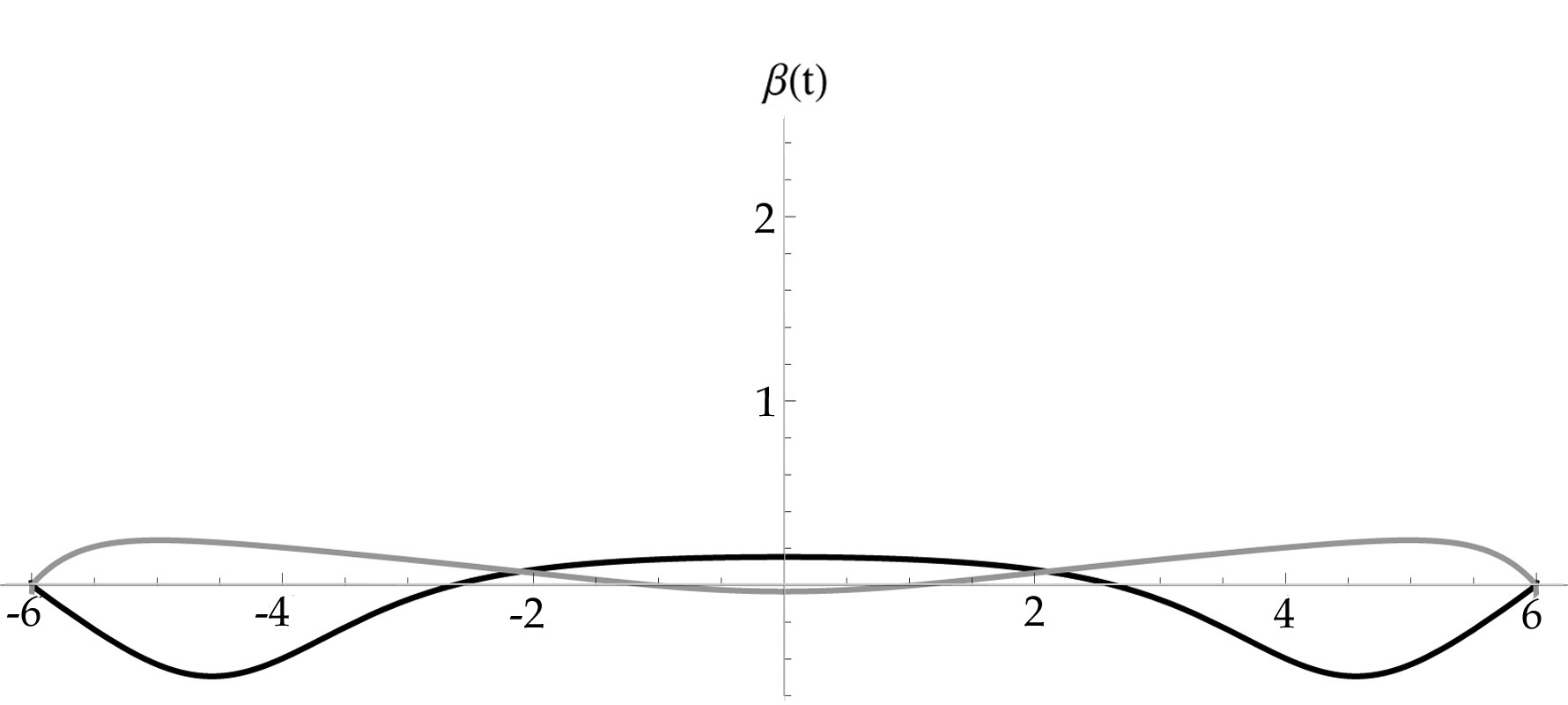}}

\end{center}
\caption{The two examples of softly varying $\beta (t)$ in  I$=[-6,6]$ for which the evolution operators imitate exactly the results of the elastic $\delta$-kicks with amplitudes $a=1$ (black line) or $a=-1$ (shadowy line). }
\label{fig:5}
\end{figure}

The above results offer some progress in describing  a class of  non-singular dynamical operations on the traditional QM level. It looks that the dynamical state transformations described above, can occur under the influence of softly varying external fields, perhaps, in standing wave equivalents of the laser beam traps? (or in some variants of the 'structured fields' \cite{Rocio}?).  Below, we shall try to see whether they can be achieved for charged particles in the traditional ion  traps.

\bigskip

\section{The quasistatic approximation}

The description of quantum states  in the ion traps depends on multiple  idealizations 
which may fail in various points. 
Thus, the commonly used oscillator potentials in the Penning and Paul  traps neglect the granular structure of the trap walls (which are never exactly smooth!). Moreover,   the  traps  are typically designed  to maintain the charged particles to investigate their internal degrees rather than to manipulate the position and momentum states. In some cases  (including the mass spectroscopy) the  trap is just  a system of  parallel metal bars \cite{LineTraps}, so  the charged particle can be efficiently trapped  without obeying the Mathieu equations. Moreover, even for almost perfect hyperbolic walls,  there are still multiple  problems concerning the quadratic Hamiltonians \eqref{n1}.

The most important  ones are the limitations of  non-relativistic approach. In reality, any potential pulse applied to the trap surface  needs some time to propagate over the trap walls and to penetrate into its  interior. Moreover, even if the delay is negligible, the too fast field changes can  awake the radiative pollution, usually neglected in the trap descriptions.

 The cases of dynamical transformations generated with some  precision are indeed not many.  The results for  the spin (qubit) states seem promising  \cite{Gisin,Gisin2,Zeilinger,Campbell} though usually limited to finite dimensional Hilbert spaces.  A notable success was achieved by  techniques of inducing the Rabi rotations between the pairs of bound states of hydrogen-like atoms (the "pink states" \cite{Harotalk}). Nonetheless, the paradoxes of quantum mechsanics are  still showing some intriguing possibilities \cite{Leggett}.   A progress is also noticed in "quantum tomography" \cite{Mancini} (see, e.g.,   the "robust" results in \cite{Marmo}). Our algorithms of Sec.7, while simplifying a part of the  problem, are still far from this level. Yet, they may suggest a  new advance in the traditional domain.

Instead of just  keeping the ions to monitor their internal structure, it could be as interesting to inject a single charged particle  into the variable trap fields, and manipulate its position-momentum 
state to check with some more details  the traditional  motion laws of quantum mechanics. The fields in the trap should be coherent, without inducing the  sharp absorption/emission effects (i.e., slowly varying, long waves, formed by clouds of tiny quanta). The laser cooling \cite{Dehmelt} could help  at the preliminary stage, but not during the proper evolution experiments, which should  neither be interrupted   ('decoupled'?) by sudden kicks \cite{Viola,Knill}. To avoid the packet  reflections 
from the walls, the trap should be wider than the traditional $r_0=5mm$ (see Paul \cite{Paul}).  But how ample could it be? The answer  depends on the relativistic corrections.  
  
In the relativistic theories a useful tool is the traditional Einstein-Infeld-Hofmann (EIH) method \cite{EIH} of develloping the physical data into the increasing powers of $\frac{1}{c}$ (Newtonian, post-Newtonian, etc.). In our case, if the interval $\text{\footnotesize [ -T},\text{\footnotesize T]}$ in Propositions 3-7 is too short, the functions $\theta$ and $\beta$ oscillate fast; so the  EIH terms will show the insufficiency of the quadratic formula \eqref{n1}. However, an inverse option of using soft and slowly changing fields in longer time intervals might reduce the  difficulty. 

 Indeed, imagine that the trap surface $\Sigma$ is a perfectly conducting (metallic) shell, composed of a number of disjoint 
leafs,  $\Sigma=\Sigma_1\cup...\cup \Sigma_n$,  surrounding a certain operation domain $\Omega\subset\mathbb{R}^3$. Consider then a static  charge distribution $\rho$ on $\Sigma$ creating a scalar potential $\phi$, constant on every connected leaf $\Sigma_j\subset \Sigma$ and harmonic in  $\Omega$, corresponding to  a solution of the traditional electrostatic problem. Assume now, that the charge density on $\Sigma$ starts to evolve  as   $\rho({\bf x})\rightarrow\rho({\bf x},t)=\beta(t)\rho({\bf x})$, where $\beta(t)$ is a certain continuous field amplitude with  $\beta(0)=1$. If such a homogeneous change of the charge density could be produced all over $\Sigma$, then, in agreement with the linearity of the theory, the local change of the scalar potential $\phi$ on $\Sigma$ would be  $\phi ({\bf x'})\rightarrow\phi({\bf x'},t)=\beta(t)\phi({\bf x'})$. However, what will be the response of the rest of the field in $\Omega$?  

In the Coulomb gauge it would be just an instantaneous effect of the changing surface density $\rho({\bf x},t)$, given by: $\int_\Sigma \frac{\rho({\bf x'},t)}{|{\bf x}-{\bf x'}|}d_2{\bf x'}=\int_\Sigma \frac{\beta(t)\rho({\bf x'})}{|{\bf x}-{\bf x'}|}d_2{\bf x'}=\beta(t)\phi({\bf x}) $. However, the Coulomb gauge is not convenient  to describe the field propagation (it requires the non-local 'transversal currents' \cite{Jack}).  To avoid this, it is easier to work in the Lorentz gauge, where the '{\em absolute time}' $t$ in the Coulomb   integral must be    
replaced by the retarded time $t_r=t-\frac{|{\bf x}-{\bf x'}|}{c}$ integrated over all sources  \cite{Grif, Jack}. Then, one can observe  that for $\rho({\bf x},t)=\beta(t)\rho({\bf x})$ 
the $\frac{1}{c}$-contributions unexpectedly simplify, resembling the effect noticed by 
Griffiths \cite{Grif}. The amplitude $\beta$ does not even need to be analytic;  if  only: $\beta (t+h)=\beta(t)+h\beta'(t)+\frac{1}{2}h^2\beta''(t)+o(h^3)$, then $\phi({\bf x},t)$ inside 
of $\Omega$ reduces to: 

\begin{multline}
\label{scalar}
\phi({\bf x},t)=\int_\Sigma \frac{\beta(t-\frac{|{\bf x}-{\bf x'}|}{c})\rho({\bf x'})}{|{\bf x}-{\bf x'}|}d_2{\bf x'}\\ =\beta(t)\phi({\bf x})-\beta'(t)\frac{Q}{c} + \frac{1}{2}\beta''(t)\frac{1}{c^2}\int_\Sigma|{\bf x}-{\bf x'}|\rho({\bf x'})d_2{\bf x'}+...
\end{multline}
where $Q$ is the initial static  charge on  $\Sigma$ and the 
last term represents the first non-trivial correction of  the delay mechanism. In addition,  if the operation takes place in a traditional ion trap, where $\Sigma$ splits into several disjoint leafs  with opposite charges, then the  $\frac{1}{c}$-terms  completely cancel and the last term of \eqref{scalar} is the only $\phi$-correction of the $\frac{1}{c^2}$ EIH-order. 

 In Paul's description, the time dependence was periodic and the dimensionless time coordinate was $\frac{\omega t}{2}$. If, however, the time dependence of $\beta(t)$ is arbitrary, then it is convenient to describe the changes of $\rho$ on $\Sigma$  in terms of a fixed time unit $\tau$ independent of any frequency.  It can be done  by replacing $t_r\rightarrow t_r/\tau$ leading to the correction $\delta\phi= \frac{1}{2}\beta''(\frac{t}{\tau})\frac{1}{(c\tau)^2}\int_\Sigma|{\bf x}-{\bf x'}|\rho({\bf x'})d_2{\bf x'}+...$, which can be also written as 
\begin{equation}
\label{correction}
\delta\phi= \frac{1}{2}\beta''(\frac{t}{\tau})\frac{1}{(c\tau)^2}\int_\Sigma\frac{|{\bf x}-{\bf x'}|^2\rho({\bf x'})}{|{\bf x}-{\bf x'}|}d_2{\bf x'}= \frac{1}{2}\beta''(\frac{t}{\tau})\int_\Sigma\frac{\varepsilon({\bf x},{\bf x'})\rho({\bf x'})}{|{\bf x}-{\bf x'}|}d_2{\bf x'}
\end{equation}

where the dimensionless amplitude in the integral of \eqref{correction} is:

 \begin{equation}
\label{epsilon}
\varepsilon({\bf x},{\bf x'})= \frac{|{\bf x}-{\bf x'}|^2}{(c\tau)^2}
\end{equation}

The real size of the trap is of course not infinite, but it seems that to perform the operations 
of our Sec.7 it can be much wider than $r_0\simeq 5mm$ assumed in the Paul's report, specially, if the 
the time unit $\tau$ is not too small. In fact, by assuming $\tau=10^{-2}s$ then $r_0=5cm$, and  
the distances $|{\bf x}-{\bf x'}|$ limited by $R=30cm$ in the real trap laboratory, we would 
end up with a tolerant estimation

 \begin{equation}
\label{eps}
\varepsilon({\bf x},{\bf x'})\leq \frac{R^2}{(c\tau)^2}\simeq 10^{-14}
\end{equation}

Apart of the scalar potential $\phi({\bf x},t)$ what matters is  the vector potential ${\bf A}({\bf x},t)$ 
created by the external currents  ${\bf j}_{ext}({\bf x},t)$ needed to feed the variable charge density ${\bf \rho}({\bf x},t)$.  What we  know about them  is only that when arriving at the surface $\Sigma$, they must obey 
 ${\bf j}_{ext}({\bf x},t)= \beta'(t) {\bf j}_{ext}({\bf x})$, to assure the charge density accumulating as 
 $\beta (t) {\bf \rho}({\bf x})$.    The vector potential  ${\bf A}({\bf x},t)$   enters into the dynamical equations in two places: (1) by contributing to the expression for  the electric field, ${\bf E}= -\nabla \phi -\frac{1}{c}\frac{\partial{\bf A}}{\partial t}$ and (2) by defining the magnetic field ${\bf H}= \nabla\times {\bf A}$ (both depending on the geometry of the external currents, but contributing to the motion equations with  $\frac{1}{c^2}$ - EIH terms only). 

In case of cylindric traps with variable potentials   $\phi({\bf x},t)= \beta(t) (\frac{x^2}{2}-\frac{y^2}{2})$, one might imagine a particle crossing the trap in the $z$-direction with the velocity $v_z=\frac{p_z}{m}$  defined well enough to feel the influence of $\beta(t)$ in some given operation interval, with the perpendicular momenta $p_x, p_y$ not too high, so that the partial states on the $x,y$- plane would perform the evolution processes described by the quadratic Hamiltonians \eqref{n1}.\footnote{Comparatively, in order to understand why the pulsating, homogeneous electric fields  permit to predict so exactly  the {\em Rabi rotations} one has to remember that a typical hydrogen-like atom (e.g. Rubidium) is of the size of 1~\AA, 
while the red light wave has the longitude of 7000~\AA; so the irradiated atom "does not see" the  space dependence of the light wave, it just reacts to the homogeneous, oscillating electric field, causing the observed Rabi effect.}

The exact control of the initial and final operation moments and of the initial and final particle state are still absent,  so our description is an intuitive rather than empirical design. What it implies, however, is that all previously described operations can indeed occur in the trap interiors, thus  confirming the reality the squeezed Fourier,  the  retarded, advanced or inverted  free evolution incidents, as well as the simulation of the positive or negative pulses of the oscillator potential \cite{BM77,Loops,Control,DavInt,Royer,David92,Dav11}.  
\bigskip

The unsolved  problems, however, are two.  (1) how  to produce the  pulse amplitudes $\beta(t)$ depending arbitrarily on time?  (2) how to assure that they will appear at the points of the trap surface simultaneously? Should one imagine a dense net of cables running from a common source of voltage to the net of points on the surface $\Sigma$? 
Should these technologies be achieved (or at least approximated), can it be expected  that the ion traps of the internal diameters 5$cm$ (or more) instead of 5$mm$ could become the efficient "wave packet laboratories"?

\section{Fundamental problems.}

Appart of purely technical challenge,  the quantum state manipulations can give us a hint 
about the validity limits of our  theories, including the controversies about fundamental 
problems  such as the time-energy uncertainty, which might still deserve  some comments.  
\bigskip

{\em Does it exist the "time operator"? }

One of reasons which could support the time-energy uncertainty principle was the idea that the  {\em time} and {\em energy} are a pair of canonically conjugate space-time  observables ($x^0$, and $p_0$)  whose quantum equivalents should therefore obey    $[{\bf t}, {\bf E}]= i\hbar\Rightarrow \Delta {\bf t} \Delta {\bf E}\ge \frac{\hbar}{2}$.
The idea seemed verbally plausible \cite{Screen} and the collection of difficulties was not immediately noticed.  Yet, it was never obvious why the probability of the "time of arrival"  should be normalized in time (the particle may arrive at the detector many times or never). The most dangerous paradox, though,  was the {\em Theorem of Pauli}.  
If the hypothetical  time and energy operators  ${\bf t}, {\bf E}$ are self-adjoint and fulfill  $[{\bf t}, {\bf E}]= i\hbar$, then both must have continuous, translation invariant spectra.  In particular, $ {\bf E}$ cannot have the lower bound, against all known facts concerning the energy operators! 

\bigskip

In an  effort to avoid the Pauli's paradox, Kijowski \cite{Kij} considered the 1-dimensional simplified model  replacing  the free energy operator 
 $ {\bf E}=\frac{{\bf p}^2}{2}$  by ${\bf \tilde{E}}=\frac{{\bf p}|{\bf p}|}{2}$ with an infinite spectrum covering $\mathbb{R}$, and then constructed the canonically conjugated 'time' 
${\bf \tilde{t}}$ (an idea followed implicitly in a trend of papers \cite{Tate,Muga2000}). However, the physical interpretation  turned artificial. In order to avoid troubles, the formalism should be limited either to the wave packets localized to the left from the detector and  moving to the right (the 'right movers'), or vice versa. Yet, to restrict the problem to the 'left component' is not the same as to consider the 'right movers' (and vice versa). Moreover,  the obtained probability distribution excluded the interference of both (left and right) components at the detection point  against the basic principles of quantum theory (the difficulty discovered by Leavens  \cite{Leav98,MuLe,Leav02, Leav05, LeavReply,Gabino,Przan}). In spite of subsequent works on POV measures and some new but still partial results \cite{Galap,Galapon,Gabino2}, it turns obvious that the idea of time-operator has too many  gaps  to grant the {\em time-energy} uncertainty (except if some additional limitations of  
quantum measurements proved real \cite{Gabino3}).  

\bigskip
Meanwhile, the counter-arguments of Aharonov, Bohm and other authors \cite{Aha61,Aha1990,Aha2000,Aha2002} grow stronger. In terms of our proposals, it is not even necessary to assure that the transformation $p\rightarrow \tilde{q}$ (squeezed Fourier) of our Sec.7  can be performed arbitrarily fast. Enough,  
if it will be achieved with sufficient precision after some finite time (call it $2\text{\footnotesize T}$). If 
the new particle position $\tilde q$ will admit a measurement precise enough,  then the measurement of the initial (free) energy $E=\frac{p^2}{2m}=\frac{1}{b^2} \frac{{\tilde{q}^2}}{2m}$, can turn too exact 
for the {\em time-energy} uncertainty (following the suggestion in \cite{Aha2002}). Though the practical solution is still missing, what matters are not the technical difficulties, but the absence of  any universal barriere. So, is the barrier indeed absent?    
\bigskip
 
{\em Is there a minimal distance?} The techniques described in Sec.7 could  provide also some insight into the quantum structure 
"in the little", e.g. to check some ideas about the universal limits which could forbid too small distances or too great momenta \cite{Gross,Witten}. It includes, in particular, the hypothesis that the position measurements by Heisenberg microscope cannot shrink the wave packets beyond some minimal size, related to the Planck distance \cite{Doplich}. However, one has to remember that  the arbitrarily narrow wave packets are just the linear combinations of much wider ones  (the Fourier transforms, wavelets, etc.). So, if the evolution laws are indeed linear (the discussions seem  not yet concluded; see  \cite{Gisin} but also \cite{Brody}) and if only the most insignificant squeezing of all position variables ${\bf q}\rightarrow \kappa {\bf q}$, ${\bf p}\rightarrow\frac{1}{\kappa}{\bf p}$, $|\kappa|<1$ could be performed (and, of course repeated) it would violate any limits of the wave packets compression and similarly, 
abolish any upper bound of the packet momentum. Hence, if such limits exist, they should be indirectly observed, by failures of the squeezed Fourier, or ordinary coordinate squeezing even in mesoscopic levels. 

Quite similar remarks concern the hypothesis of non-commutative geometries of the space coordinates (assumed sometimes to prevent the singularity creation below the Planck scale \cite{Madore,Witten}). The simplest case of $[x,y]=i\nu$, (with $0\neq\nu\in \mathbb{R}$), if $x$ and $y$ were interpreted as the coordinates of some microobject, could not survive even the most insignificant squeezing of both $x$ and $y$.   

Some less destructive implications might still exist without challenging the spacetime structure. Indeed, if 
the squeezing/amplification can be generated for the continuous space variables, then it might help to 
observe some traditional quantum effects. In fact, the quantum mechanical particle-wave duality is difficult to observe for beams of heavy 
particles, with very narrow wave fronts running almost along the classical trajectories. If, however, the 
coordinate amplification can be engineered, expanding the wave fronts  without affecting the coherence, then  it could help to detect the heavy particles interference, or else, the duality limitations if they exist.  

Appart of these particular problems, the  techniques of the dynamical state transformations might still check  some more general  mysteries. In spite of their successes, it seems frustrating that all quantum theories were constructed 
just by multiplying {\em ad infinitum} the same linear state-observable structure, while leaving the basic paradoxes  almost forgotten. Until now, the scheme  never failed.   
 But, how ample is indeed the orthodox truth?
\bigskip

{\em Acknowledgements} The author is indebted  to David Fernandez, Piotr Kielanowski, Francisco Delgado and Alonso Contreras for helpful comments,  to Jesus Fuentes Aguilar for technical assistance. The support of the CONACYT project 152574 is acknowledged. 

\end{document}